\documentclass[useAMS,usenatbib]{mn2e}
\usepackage{epsfig}
\usepackage{times}
\usepackage{mathptm}
\usepackage{color}

\newcommand{\dd}{{\rmn d}}

\title[Searching for occultations in young open clusters] {The
  \emph{Monitor} project: Searching for occultations in young open
  clusters}

\author[S.~Aigrain et al.]
 {S.~Aigrain$^1$\thanks{E-mail: suz@ast.cam.ac.uk}, S.~Hodgkin$^1$, 
  J.~Irwin$^1$, L.~Hebb$^2$, M.~Irwin$^1$, F.~Favata$^3$, \newauthor
  E.~Moraux$^4$, F.~Pont$^5$ \\
  $^1$Institute of Astronomy, University of Cambridge, Madingley
  Road, Cambridge, CB3 0HA, United Kingdom \\
  $^2$School of Physics and Astronomy, University of St Andrews,
  North Haugh, St Andrews KY16 9SS, Scotland \\
  $^3$ESA/ESTEC, Keplerlaan 1, PO Box 299, 2200 AG Noordwijk, 
  The Netherlands \\
  $^4$Laboratoire d'Astrophysique, Observatoire de Grenoble, BP 53,
  F-38041 Grenoble C{\' e}dex 9, France \\
  $^5$Observatoire Astronomique de l'Universit{\' e} de Gen{\` e}ve, 51, 
  chemin des Maillettes, CH-1290 Sauverny, Switzerland}

\begin{document}

\date{Accepted \ldots Received \ldots; in original form \ldots}

\pagerange{\pageref{firstpage}--\pageref{lastpage}} \pubyear{2002}

\maketitle

\label{firstpage}

\begin{abstract}

  The \emph{Monitor} project is a photometric monitoring survey of
  nine young ($1$--$200$\,Myr) clusters in the solar neighbourhood to
  search for eclipses by very low mass stars and brown dwarfs and for
  planetary transits in the light curves of cluster members. It began
  in the autumn of 2004 and uses several 2 to 4\,m telescopes
  worldwide. We aim to calibrate the relation between age, mass,
  radius and where possible luminosity, from the K-dwarf to the planet
  regime, in an age range where constraints on evolutionary models are
  currently very scarce. Any detection of an exoplanet in one of our
  youngest targets ($\la 10$\,Myr) would also provide important
  constraints on planet formation and migration timescales and their
  relation to proto-planetary disc lifetimes. Finally, we will use the
  light curves of cluster members to study rotation and flaring in
  low-mass pre-main sequence stars.

  The present paper details the motivation, science goals and
  observing strategy of the survey. We present a method to estimate
  the sensitivity and number of detections expected in each cluster,
  using a simple semi-analytic approach which takes into account the
  characteristics of the cluster and photometric observations, using
  (tunable) best-guess assumptions for the incidence and parameter
  distribution of putative companions, and we incorporate the limits
  imposed by radial velocity follow-up from medium and large
  telescopes. We use these calculations to show that the survey as a
  whole can be expected to detect over 100 young low and very low mass
  eclipsing binaries, and $\sim 3$ transiting planets with radial
  velocity signatures detectable with currently available facilities.

\end{abstract}


\section{Introduction}
\label{sec:intro}

Mass is the most fundamental property of a star, yet direct
measurements of stellar masses are both difficult and rare, as are
measurements of stellar radii. Detached eclipsing binary systems
provide the most accurate determinations (to $\sim 2\%$) of the mass
\emph{and} radius of both components \citep{and91}, which are
(reasonably) assumed to have a single age and metallicity. These
systems therefore provide extremely stringent tests of stellar and
sub-stellar evolutionary models. Temperatures and distance independent
luminosities, which are also needed to constrain the models, are also
derived from the analysis of eclipsing systems. If the companion is
too faint to allow the detection of a second set of lines in the
spectrum or of secondary eclipses, useful measurements of the mass and
radius ratios of co-eval systems can still be obtained. Large numbers
of eclipsing binaries are now known in the field, and evolutionary
models are thus relatively well-constrained on the main sequence,
though more discoveries of very-low mass eclipsing binaries would be
desirable. However, very few such systems have yet been discovered in
open clusters, allowing a precise age measurement, and even fewer in
young (pre-main sequence) clusters and star-forming associations,
giving constrains on the crucial early stages of stellar and
sub-stellar evolution.
  
Similarly, transiting extra-solar planets are particularly interesting
because both their radius and their mass can be measured (relative to
their parent star, using photometry and radial velocity measurements),
giving an estimate of their density and hence of their composition. At
the present time, a handful of planets are known to transit their
parent stars, but all are in the field, and there are no radius
measurements of young planets.

Photometric monitoring of young open clusters is the only way to
systematically search for young occulting systems with well known ages
and metallicities. \citet{hwg04} have demonstrated that this technique
can be used to probe down to low masses in older open
clusters. Monitor aims to reach even lower masses at younger ages. The
most fundamental result expected from the Monitor project as a whole
is the calibration of the mass-radius relation from M-stars to
planets, throughout the pre-main sequence age range. In the next
section, we examine existing constraints on this relation.

\subsection{Existing constraints on the mass-radius relation}
\label{sec:mrrel_constraints}

Constraints on the mass-radius relation at early ages are the most
fundamental science outcome expected from the Monitor project. This is
because, aside from a small number of bright objects with known masses
and distances whose radius can be measured interferometrically,
detached double-lined eclipsing binaries provide the tightest, and the
only model-independent constraints on the masses and radii of stars
and brown dwarfs, and transiting planets are the only ones for which
we can measure radii at all.

\begin{figure}
\centering
\epsfig{file=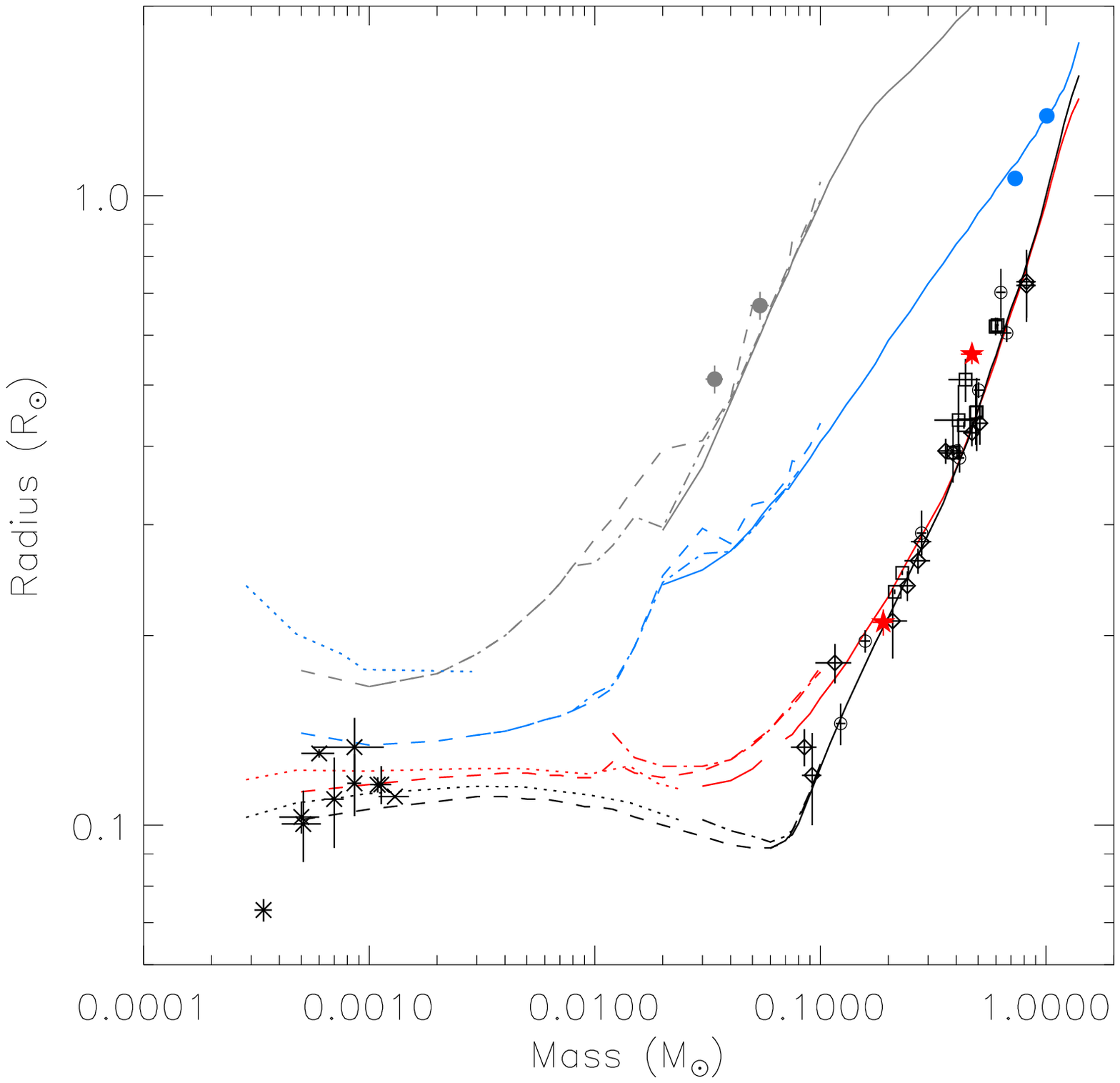,width=\linewidth}
\caption{Observational constraints on the mass-radius relation.  Black
  circles represent interferometric measurements of field stars
  \protect\citep{lbk01,skf+03}, and all other symbols represent
  members of eclipsing binary or transiting systems: the secondaries
  of field F-M or G-M systems from the OGLE survey
  \protect\citep{bpm+05,pbm+05} are shown as black diamonds, field M-M
  systems \protect\citep{mml+96,tr02,rib03,mm04,cbb+05,lr05} as black
  squares, and planets that transit across field stars
  (\protect\citealt{cbl+00,hmb+00,ktj+03,bps+04,kts+04};
  \protect\citealt{pbq+04,abt+04,kts+05,bum+05,sfh+05,msv+06}) as black
  crosses. The red filled stars represent the NGC\,1647 system
  \protect\citep{hwg+06}, the blue filled circles the Ori\,1c
  system \protect\citep{smv+04} and the grey filled circles the ONC
  double brown dwarf system \protect\citep{smv06}. The solid, dashed,
  dot-dash and dotted lines respectively show the NEXTGEN
  \protect\citep{bca+98}, DUSTY \protect\citep{cba+00} and COND
  \protect\citep{bcb+03} models of the Lyon group and the non-grey
  models of \protect\citet{bmh+97} for 1\,Gyr (black), 150\,Myr (red),
  10\,Myr (blue) and 1\,Myr (grey).
  \label{fig:mrrel_constraints}}
\end{figure}

In recent years, the discovery of a number of eclipsing binaries with
at least one M-star component
\citep{tr02,rib03,mm04,cbb+05,lr05,bpm+05,pbm+05} together with
interferometric radius measurements of a number of field dwarfs in the
late-K to mid-M spectral range \citep{lbk01,skf+03}, have vastly
improved the available constraints on the low-mass main sequence
mass-radius relation, down to the very edge of the brown dwarf regime
\citep{pmb+05,pmb+06}. These form a tight sequence which is relatively
well reproduced by evolutionary models of low-mass stars such as those
of \citet{bca+98}, as illustrated in
Figure~\ref{fig:mrrel_constraints}.

On the other side of the `brown dwarf desert', the ten planets that
are currently known to transit their parent star
(\citealt{cbl+00,hmb+00,ktj+03,bps+04,kts+04};
\citealt{pbq+04,abt+04,kts+05,bum+05,sfh+05,msv+06}) present very
diverse properties even at late ages, some falling above or below the
locus predicted by models of isolated gaseous objects without a solid
core \citep{bmh+97,bcb+03}. Evolutionary models incorporating solid
cores and the effects of tidal interaction with and irradiation by the
parent star are now successfully reproducing the radii of most of
them, except for HD\,209458b (the most massive of the group of two
large planets on Figure~\ref{fig:mrrel_constraints}), which remains a
challenge \citep{lwv+05,bcb+05}.

However, only six data points so far constrain the mass-radius
relation from $1\,M_{\sun}$ downwards at ages younger than 1\,Gyr. The
first pair is a $1.0+0.7\,M_{\sun}$ eclipsing binary discovered by
\citet{smv+04}, thought to belong to the Ori\,1c association and with
an estimated age of 5--10\,Myr (shown in blue on
Figure~\ref{fig:mrrel_constraints}). The next is a double M-star
eclipsing binary found by \citet{hwg+06} in the 150\,Myr old open
cluster NGC\,1647 (shown in red on
Figure~\ref{fig:mrrel_constraints}). Finally, the recent discovery by
\citet{smv06} of a double brown dwarf eclipsing binary in the
$\sim1$\,Myr Orion Nebula Cluster (shown in grey on
Figure~\ref{fig:mrrel_constraints}) represents, to the best of our
knowledge, the first direct constraint on the mass-radius relation for
brown dwarfs at any age. All of these objects fall significantly above
the main-sequence relation, highlighting the importance of age in this
diagram.

\citet{smv+04} and \citet{hwg+06} compared the properties of the first
two EBs to a number of evolutionary models in the literature -- some
of the most widely used are illustrated on
Figure~\ref{fig:mrrel_constraints} -- but none were found that fit
both components of each system simultaneously (although the
discrepancy is not clearly visible on
Figure~\ref{fig:mrrel_constraints}, the relevant isochrone
systematically misses the error box on at least one of the components,
a problem which is not solved either by adjusting the age or using a
different set of isochrones). While the masses and radii of the latter
ONC EB are in reasonable agreement with theoretical models for the
assumed age of the system, the (less massive, fainter) secondary
appears to be hotter than the primary. None of the models predict this
surprising result.

Monitor has been designed to attempt to populate the entire section of the
diagram in Figure~\ref{fig:mrrel_constraints} which lies above the
main sequence line, across the entire range of masses shown.

\subsection{Young low-mass binaries}
\label{sec:young_binaries}

Aside from improving our understanding of the mass-radius relation,
simply measuring dynamical masses for stars and brown dwarfs with
known distances and ages provide important constraints on the
evolutionary models of these objects. Dynamical masses can be
obtained by spectroscopic follow-up of eclipsing binaries or by direct
searches for spectroscopic binaries, which is foreseen in those
clusters that lend themselves to it as an extension of the main,
photometric part of the Monitor project.

The distribution of stellar masses is a direct result of the star
formation process. Measurements of individual stellar and sub-stellar
masses provide crucial information on the structure and evolution of
these objects \citep{lv02} and measurements of the mass function of a
population allow us to understand the detailed physics of the star
formation mechanism as a whole.
  
The stellar mass function has long been studied \citep[e.g.][]{sal55},
but only recently are we pushing down into the low-mass and
sub-stellar regimes
\citep{hc00,bbs+02,lbs+03,mbs+03,shc04}. Luminosity functions can be
reliably determined, but the more fundamental mass functions remain
uncertain, particularly at the low-mass end and for young ages. In
this regime, conversion of a luminosity function to a mass function
relies heavily on theoretical stellar evolutionary models which infer
stellar masses and ages from derived luminosities and
temperatures. These models suffer from large uncertainties at low
masses, because of the complexity of modeling stellar atmospheres
below 3800\,K, where molecules dominate the opacity and convection
dominates energy transport \citep{hw04}, and at early ages, because of
the lack of observational constraints on the initial conditions
\citep{bca+02}.
  
Stellar and sub-stellar masses can only be determined through
investigation of an object's gravitational field, and the small subset
of objects in which mass determination is possible are used to
calibrate evolutionary models for all stars. A small (but growing)
number of low-mass main sequence stars have empirically measured
masses \citep{dfs+00,lr05}, however the models are not fully
constrained at young ages or at the lowest masses (brown dwarf regime)
due to the scarcity of measurements, which only increases towards the
brown dwarf domain \citep{bdk+04}. There are only a handful of
dynamical mass constraints for low-mass {\it PMS} objects
\citep{dfs+00,hw04,smv+04,clg+05,smv06}, and these constitute a
growing body of evidence suggesting that current evolutionary models
systematically under-predict masses of pre-main sequence stars (for a
given temperature or luminosity) below $0.5\,M_{\sun}$.

More dynamical mass measurements of young very-low-mass stars and BDs
of known age are clearly needed to anchor the theory. The only way to
do this in a systematic way is by searching for binaries in young
clusters and star forming regions. As the low-mass members of clusters
which are rich enough to provide statistically significant numbers of
targets tend to be too faint for the current capabilities of direct
imaging and astrometric searches, spectroscopy (radial velocities) and
photometry (occultations) appear to be the most promising methods.

\subsection{Planets around young stars}
\label{sec:young_planets}

The detection of young planets not only helps to anchor evolutionary
models of planets -- as constrained by the mass-radius relation -- but
also improves our understanding of the formation of planetary
systems and of the dynamical processes that take place early on in
their evolution.

The past decade has seen the discovery of nearly 200 extra-solar
planets (ESPs), mainly via the radial velocity (RV) method.
Statistical studies of these systems \citep[see
e.g.][]{ums03,sim+03,eum04} provide constraints on formation and
migration scenarios, by highlighting trends in the minimum mass-period
diagram, or in incidence rate versus parent star metallicity. However,
almost all the currently known planets orbit main sequence field stars
whose ages can be determined only approximately. Any detection of a
planet around a pre-main sequence star would provide much more direct
constraints on formation and migration timescales, particularly around
a star aged 10\,Myr or less, the timescale within which near-infrared
observations \citep{hll01} and accretion diagnostics \citep{jcs+06}
indicate that proto-planetary discs dissipate.

The only known planetary mass companion within that age range is the
companion to 2MASS 1207334--393254, a member of the $\sim8$\,Myr
association TW~Hydra \citep{cld+05}. This system's properties are more
akin to those of binaries than star-planet systems separation
\citep[see e.g.][]{ldc05}, and the detection of other, more typical
planetary systems in this age range is a major possible motivation for
the Monitor project.

\subsection{Existing open cluster transit surveys}
\label{sec:other_surveys}

The potential impact of any transit discovery in an open cluster,
together with other advantages such as the fact that an estimate of
their masses can usually be determined from their broad-band colours
alone, have motivated a number of transit searches over the last few
years. These include the UStAPS (the University of St Andrews Planet
Search, \citealt{shl+03,bhm+05,hck+05}, EXPLORE-OC \citep{vls+05},
PISCES (Planets in Stellar Clusters Extensive Search,
\citealt{mss+05,mss+06}) and STEPSS (Survey for Transiting Extrasolar
Planets in Stellar Systems, \citealt{bgd+06}). Some of these surveys
are still ongoing, but no detection of a transiting planet confirmed
by radial velocity measurements has been announced so far. These
non-detections are at least partially explained by initially
over-optimistic estimates of the detection rate, and by the smaller
than expected number of useful target stars per cluster field.

A direct comparison of Monitor to the existing surveys in terms of
observational parameters is somewhat complex, because of the range of
telescopes and strategies adopted in Monitor (see
Section~\ref{sec:survey}). However, broadly speaking, Monitor uses
similar observing cadence as previous surveys, but has to use
generally larger telescopes (2.2--4\,m rather than 1 to--2.5\,m),
because of its focus on young clusters, which limits the choice of
target distances. This implies that Monitor surveys are somewhat
deeper, but that it is not possible to obtain continuous allocations
of several tens of nights (as the telescopes in question are in heavy
demand). The number of cluster members monitored with sufficient
photometric precision to detect occultations in each cluster varies
from several hundred to over 10\,000, i.e.\ it is in some cases lower
than the typical numbers for other surveys, and in others higher.

More importantly, aside from these observational considerations, there
are fundamental differences between Monitor and other open cluster
transit surveys. The first is that it focuses on younger (pre-main
sequence) clusters (the youngest target clusters of the above surveys
are several hundred Myr old). Any detections arising from Monitor
would thus have a different set of implications to those arising from
a survey in older clusters. The second is that Monitor was designed to
target lower mass stars, because that is where constraints on
evolutionary models and companion incidence were the scarcest and
because the youth of our targets made it possible (low mass stars
being brighter at early ages, compared to their higher mass
counterparts). Finally, while the aforementioned surveys have been
designed with the explicit goal of searching for planetary transits,
the detection of eclipsing binaries was considered as important as
that of planetary transits when choosing Monitor targets and observing
strategies. As we shall see, detections of binaries are expected to
far outnumber detections of planets. Compared to other open cluster
transit surveys, Monitor thus explores a very different area of the
complex, multi-dimensional parameter space of double star and
star-planet systems.
  
\subsection{Additional science}
\label{sec:add}
  
The proposed observations are also ideally suited to measuring
rotational periods for various ages and masses. The age distribution
of our target clusters samples all important phases of the angular
momentum evolution of low-mass stars, including the T~Tau phase where
angular-momentum exchange with an accretion disk is important, the
contraction onto the zero-age main sequence, and the beginning of the
spin-down on the main sequence. The high time cadence and relatively
long baselines required by the principal science goal (the search for
occultations) implies excellent sensitivity to periods ranging from a
fraction of a day to over ten days, and longer in the case of our
`snapshot mode' observations (see Section~\ref{sec:strategy}), while
our photometric precision should allow us to measure periods right
across the M-star regime, and into the brown dwarf domain in some
cases. The rotational analysis of several of our clusters is already
complete (M34, \citealt{iah+06}) or nearing completion (NGC\,2516,
Irwin et al.\ in prep., NGC\,2362, Hodgkin et al.\ in prep.), and we
refer the interested reader to those papers for more details.

In addition, we will also use the light curves collected as part of
the Monitor project to search for and study other forms of photometric
variability in the cluster members, such as flaring, micro-flaring and
accretion related variability. At a later date, the exploitation of
the light curves of field stars falling within the field-of-view of
our observations is also foreseen, including searching for
occultations and pulsations.

\smallskip

The target selection and survey design for Monitor are described in
Section~\ref{sec:survey}. In Section~\ref{sec:ndet}, we performed a
detailed semi-empirical investigation of the number and nature of
detections expected in each target cluster. In Section~\ref{sec:foll},
we describe our follow-up strategy and incorporate the limits of
feasible radial velocity follow-up into the detection rate estimates
of the previous Section. The present status of the observations,
analysis and follow-up are briefly sketched out in
Section~\ref{sec:status}.

\section{The \emph{Monitor} photometric survey}
\label{sec:survey}

\subsection{Target selection}
\label{sec:targets}

\begin{table*} 
  \begin{minipage}{\linewidth}
    \caption{Basic properties of the Monitor target clusters. {\bf may need to change age of ic4665} }
    \label{tab:targets}
    \begin{center}
      \begin{tabular}{@{}lrrrrrrrrrrrrl}
        \hline
        Name & RA & Dec & Age & $(M-m)_0$ & $E(B-V)$ & $\Omega_{\rm C}$ & $I_{\rm BD}$ & $M_{20}$ & $N'$ & $\Omega'$ & $M_{\rm L}'$ & $M_{\rm H}'$ & Ref \\
        & $hh~mm$ & $dd~mm$ & (Myr) & (mag) & (mag) & (\sq\degr) & (mag) & ($M_{\sun}$)$^2$ & & (\sq\degr) & ($M_{\sun}$) & ($M_{\sun}$) & \\
        \hline
        ONC               & 05~35 & $-$05~23 &   1 &  8.36 & 0.05 &        $\sim 0.35$ & 16.78 & 0.02 & $1600$ & $0.5~$ & $0.1~~$ & $50~~~$ & a \\
                          &       &          &     &       &      &                    & 16.78 & 0.02 &  $500$ & $0.07$ & $0.02~$ &  $0.5~$ & b \\
        NGC\,2362         & 07~19 & $-$24~57 &   5 & 10.85 & 0.10 &        $\sim 0.15$ & 20.31 & 0.10 &  $500$ & $0.11$ & $0.11~$ &  $0.65$ & c,d \\
       $h$ \& $\chi$\,Per & 02~20 & $+$57~08 &  13 & 11.85 & 0.56 & $2\times\sim 0.05$ & 22.85 & 0.35 &  $279$ & $1.0~$ & $4.0~~$ & $15~~~$ & e \\
        IC\,4665          & 17~46 & $+$05~43 &  28 &  7.72 & 0.18 &        $\sim 4.0~$ & 19.42 & 0.55 &  $150$ & $4.0~$ & $0.02~$ &  $0.2~$ & g,h \\
        NGC\,2547         & 08~10 & $-$49~10 &  30 &  8.14 & 0.06 &        $\geq 0.85$ & 19.21 & 0.05 &  $700$ & $0.85$ & $0.035$ &  $0.9~$ & f \\
        Blanco\,1         & 00~04 & $-$29~56 &  90 &  7.07 & 0.01 &        $\geq 2.3~$ & 19.15 & 0.06 &  $300$ & $2.3~$ & $0.03~$ &  $0.6~$ & i \\
        M50               & 07~02 & $-$08~23 & 130 & 10.00 & 0.22 &        $\sim 0.19$ & 22.68 & 0.25 & $2050$ & $0.35$ & $0.05~$ &  $0.55$ & j \\
        NGC\,2516         & 07~58 & $-$60~52 & 150 &  8.44 & 0.10 &        $\geq 2~~~$ & 20.0  & 0.08 & $1200$ & $2.0~$ & $0.02~$ &  $0.2~$ & k,l \\
        M34               & 02~42 & $+$42~47 & 200 &  8.98 & 0.10 &        $\sim 0.55$ & 21.7  & 0.11 &   $89$ & $0.55$ & $0.9~~$ &  $2.5~$ & k,m \\
        \hline
      \end{tabular}
    \end{center}
    {\footnotesize {\bf Notes:} The age, distance modulus $(M-m)_0$
      and reddening $E(B-V)$ of each cluster were taken from the
      literature ($1^{\rm st}$ entry in the `Ref' column if more than
      one is present), where they were generally derived from
      isochrone fitting to the cluster sequence on optical (and in
      some cases near-IR) colour-magnitude diagrams. The cluster area
      $\Omega_{\rm C}$ is given approximately, based on the area
      covered in the reference used and whether the entire extent of
      the cluster was covered or not. The apparent magnitude $I_{\rm
        BD}$ at the Hydrogen-burning mass limit of $0.072\,M_{\sun}$
      and the mass $M_{20}$ corresponding to $I=20$ were deduced from
      the cluster ages, distance moduli and reddening values using the
      models of \protect\citet{bca+98} and the extinction law of
      \protect\citet{bm98}. The approximate number $N'$ of known or
      candidate cluster members prior to starting the Monitor project
      was also taken from the literature ($2^{\rm nd}$ entry in the
      `Ref' column if more than one is present). Two separate values
      are given for the ONC as they correspond to widely different
      mass ranges ($M_{\rm L}'$ to $M_{\rm H}'$) and spatial coverage
      ($\Omega'$). Where the reference used quoted a number of
      candidate members (generally selected from optical CMDs using
      theoretical PMS isochrones), but also gave an estimate of the
      degree of contamination by field stars, we give here the number
      of candidate members corrected for contamination. The references
      are: (a) \protect\citet{hil97}; (b) \protect\citet{hc00}; (c)
      \protect\citet{mah+01}; (d) \protect\citet{dah05}; (e)
      \protect\citet{shm02}; (f) \protect\citet{jnb+04}; (g)
      \protect\citet{man06} (h) \protect\citet{dbp+06}; (i) 
      \protect\citet{mbs+06}; (j)
      \protect\citet{kfr+03}; (k) \protect\citet{sbg+04}; (l) Moraux
      et al.\ (in prep.) (m) \protect\citet{is93}. Where 2 references
      are given on one line, the first was used for the age, distance
      and reddening and the second for the membership estimate. }
  \end{minipage} 
\end{table*} 

The initial selection criteria for our target clusters were that their
age be $\leq 200$\,Myr, that the apparent $I$-band magnitude at the
Hydrogen-burning mass limit be $\leq 21$ (this implies an
age-dependent distance limit), and that at least a few hundred PMS
cluster members could conveniently be surveyed in a single field of
one of the available wide-field optical cameras on 2 to 4\,m
telescopes (i.e.\ that the cluster should be compact and rich enough,
with a well-studied low-mass pre-main sequence population). These
criteria were initially applied to a list of open clusters and star
forming regions compiled from the literature, the WEBDA
Open Cluster database\footnote{See {\tt
    http://www.univie.ac.at/webda/}.} and several open cluster
atlases. This yielded a list of top-priority clusters fulfilling all
of the above criteria (Orion Nebula Cluster, NGC\,2362, NGC\,2547,
NGC\,2516), which was then completed with clusters which fulfilled
only some of the criteria but filled a gap in the age sequence
constituted by the original set of targets and/or had a right
ascension which was complementary to that of another cluster, allowing
them to be observed simultaneously by alternating between the two ($h$
\& $\chi$\,Per, IC\,4665, Blanco\,1, M50, M34).

Some obvious candidates were excluded because of their large angular
extent (e.g.\ $\alpha$\,Per and the Pleiades) or because they were not
rich enough (e.g. IC\,348). NGC\,2264 will be the target of a
continuous 3-week ultra-high precision monitoring program in the
framework of the additional program of the CoRoT space
mission\footnote{CoRoT is a small (30\,cm aperture) Franco-European
  space telescope due for launch in late 2006, whose primary science
  goals are asteroseismology and the detection of extra-solar planets
  around field stars via the transit method. See {\tt
    http://corot.oamp.fr/} for more details. The PI of the CoRoT
  additional program on NGC\,2264 is F.\ Favata.}, which will far
outstrip the time sampling and photometric precision achievable from
the ground, and was therefore left out of the present survey. Two
clusters, NGC\,6231 and Trumpler\,24, appeared to be promising targets
but had poorly studied low mass populations, and a preliminary
single-epoch multi-band survey was undertaken to investigate their
low-mass memberships. Depending on the results of this survey, these
two clusters may be added to the list of Monitor targets.

\begin{figure}
  \centering
  \epsfig{file=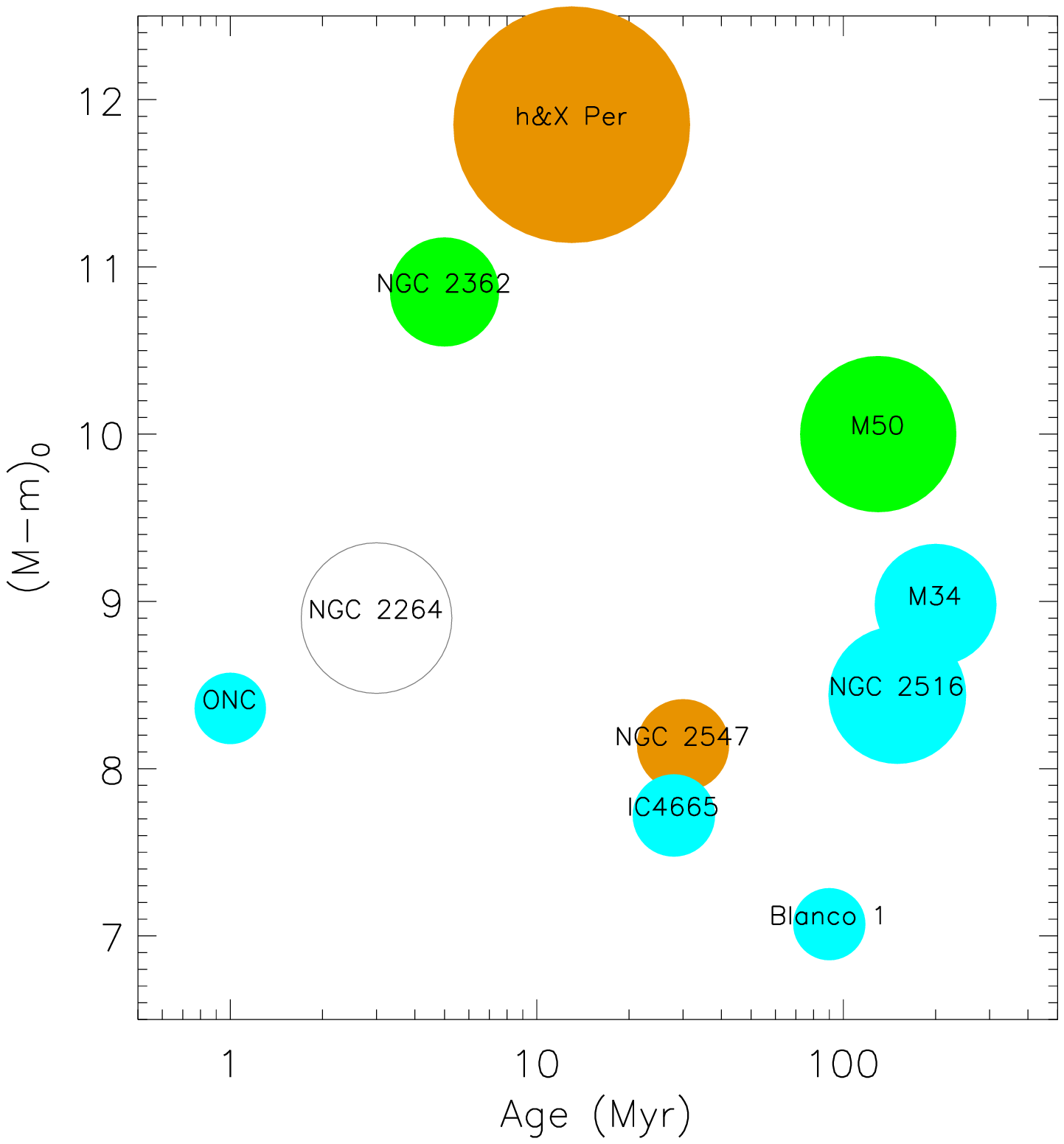,width=\linewidth}
  \caption{Age and distance distribution of the Monitor target
    clusters. The size of the circle representing each cluster scales
    with $N^{0.3}$, where $N$ is the number of non-saturated cluster 
    members monitored with better than 5\% precision (see 
    Section~\ref{sec:nsys}), and the colour-coding indicates whether
    we have obtained all (green), some (blue) or none (orange) of we
    have obtained all (green), some (blue) or none (orange) of the
    data for that cluster at the time of writing. For comparison, we
    also show NGC,2264 (hollow circle), which will be the target of a
    CoRoT monitoring campaign. 
    \label{fig:targets}}
\end{figure}

The most up to date estimates of the properties (age, distance,
reddening, membership) of the current set of target clusters that were
found in the literature are summarised in
Table~\ref{tab:targets}. Figure~\ref{fig:targets} summarises the age
and distance distributions of the target clusters together with the
number of objects monitored in each and the degree of completion of
the monitoring to date.

\subsection{Observing strategy}
\label{sec:strategy}

\begin{table*}
  \begin{minipage}{\linewidth}
    \caption{Observations of the Monitor targets to date.} 
    \label{tab:obs}
    \begin{center}
      \begin{tabular}[c]{@{}llrrrrrrrrrrcl}
        \hline
        Name & Tel & FOV & $N_{\rm P}$ & $t_{\rm exp}$ & $\delta t$ & $I_{\rm sat}$ & $I_{5\%}$ & $t_{\rm req}$ & $t_{\rm all}$ & $t_{\rm tot}$ & $T$ & filter & semester/ \\
        & & ({\sq\degr}) & & (sec.) & (min) & (mag) & (mag) & & & (h) & (days) & & period \\
        \hline
        ONC               & INT  & $0.29$ & 1 & 30  &  $3.5$ & 13.0 & 19.0 &  40\,n &  40\,n & 55.5 &  70 & $V,i$ & 04B--06B \\
        NGC\,2362         & CTIO & $0.38$ & 1 & 75  &  $6.6$ & 15.5 & 20.5 &  14\,n &  14\,n & 93.6 & 360 &   $i$ & 05A--06A\\ 
       $h$ \& $\chi$\,Per & CFHT & $1.0~$ & 1 & 120 & $10~~$ & 15.5 & 20.5 & 100\,h &  40\,h & 0.0  & --- &   $I$ & 05B,06B \\
                          & KPNO & $0.35$ & 2 & 75  &  $9~~$ & 15.5 & 20.5 &   8\,n &   8\,n & 0.0  & --- &   $i$ & 06B \\
        IC\,4665          & CFHT & $1.0~$ & 4 & 120 & $14~~$ & 15.5 & 20.5 & 100\,h &  40\,h & 16.1 & 136 &   $I$ & 05A \\
        NGC\,2547         & 2p2  & $0.29$ & 2 & 120 & $15~~$ & 13.0 & 19.5 & 100\,h & 100\,h & 0.0  & --- &   $I$ & P75--P77 \\
        Blanco\,1         & 2p2  & $0.29$ & 3 & 120 & $15~~$ & 13.0 & 19.5 & 100\,h & 100\,h &  6.6 &  26 &   $I$ & P75--P78 \\
        M50               & CTIO & $0.38$ & 1 & 75  &  $6.6$ & 15.5 & 20.5 &  14\,n &  14\,n & 93.6 & 360 &   $i$ & 05A--06A \\ 
        NGC\,2516         & CTIO & $0.38$ & 3 & 75  &  $8.8$ & 15.5 & 20.5 &   8\,n &   8\,n & 68.6 & 400 &   $i$ & 06A \\ 
        M34               & INT  & $0.29$ & 1 & 30  &  $3.5$ & 13.0 & 19.0 &  10\,n &  10\,n & 18.0 & 10  & $V,i$ & 04B \\
                          & CFHT & $1.0~$ & 1 & 120 & $10~~$ & 15.5 & 20.5 & 100\,h &  40\,h & 0.0  & --- &   $I$ & 05B,06B \\
                          & KPNO & $0.35$ & 1 & 75  &  $9~~$ & 15.5 & 20.5 &   8\,n &   8\,n & 0.0  & --- &   $I$ & 06B \\
        \hline
      \end{tabular}
    \end{center}
    {\footnotesize{\bf Notes:} The different telescope/instrumentation
      combinations used are: INT: Isaac Newton Telescope (2.5\,m) with
      the Wide Field Camera; CTIO: Blanco telescope (4\,m) at Cerro
      Tololo Interamerican Observatory with MosaicII; CFHT:
      Canada-France-Hawaii Telescope (3.6\,m) with MegaCAM; KPNO:
      Mayall telescope (4\,m) at Kitt Peak National Observatory with
      Mosaic; 2p2: ESO/MPI telescope (2.2\,m) at La Silla Observatory
      with the Wide Field Imager.  $N_{\rm P}$ refers to the number of
      pointings used for each cluster, $t_{\rm exp}$ to exposure time
      used for each pointing, and $\delta t$ to the resulting interval
      between consecutive observations. $I_{\rm sat}$ and $I_{5\%}$
      are the approximate $I$-band magnitudes at which saturation
      occurs and the frame-to-frame rms of the light curves is $\sim
      5$\%, respectively. The total observing time requested ($t_{\rm
        req}$) and allocated ($t_{\rm al}$) to date for each cluster
      are given in nights for visitor mode observations and hours for
      snapshot or service mode observations. The total time on target
      $t_{\rm tot}$, which is equal to the duration of the light
      curves produced so far with the daily gaps removed, is given in
      hours in all cases (and should be compared with the target of
      $\sim 100$\,h). $T$ refers to the total time span of the light
      curves to date. The completion rate of our INT runs to date has
      been very partial due to adverse weather conditions. $h$ \&
      $\chi$\,Per and M34 are targeted with two and three different
      telescopes respectively, covering different magnitude
      ranges. The semester/period column refers to the telescope time
      allocation semester or period (P75 corresponds approximately to
      05B).}
  \end{minipage}
\end{table*}

The optimal observing strategy for a survey like Monitor is a complex
combination of a large number of considerations including photometric
precision, number of objects monitored in a given mass range, and time
sampling. The different ages, physical sizes and distances of our
target clusters, as well as their positions on the sky, also come into
play, as do considerations of a more practical nature, such as the
need to find clusters of compatible right ascension to observe in one
given run or cycle. In general, the size of telescope to use for a
given cluster was determined by the magnitude of the Hydrogen burning
mass limit inferred from the cluster age and distance, given that we
wished to monitor objects near this limit with precision sufficient to
detect occultations, i.e\ a precision of a few percent at worst.  The
nearest and brightest clusters such as the ONC are suitable targets
for 2\,m class telescopes, whereas the more distant or older clusters
are more suitable for 4\,m class telescopes. While some attempt was
made at matching detector field-of-view (FOV) to cluster angular size,
this was not always possible -- there is currently no equivalent of the
1\,sq.deg.\ FOV of CFHT/MegaCAM in the South, so that our large
Southern targets were monitored using a dither pattern of 3 or 4
pointings.

Exposure times were adjusted to ensure a precision of 1\% or better
down to the cluster Hydrogen burning mass limit (HBML) or the apparent
magnitude $I=19$, whichever was the brightest, with the caveat that
exposures were kept sufficiently long to avoid being excessively
overhead-dominated. The $I\leq 19$ limit arises from the need to
perform radial velocity follow-up of all candidates to determine
companion masses, which becomes impractical even with 8\,m class
telescopes beyond that limit.

Adequate sampling of the event is vital to ensure the detection is of
an occultation and not of some other type of temporary dip in flux. In
conventional planetary transit searches around field stars, the shape
of the candidate transit event is used to minimise contamination by
stellar eclipses. In the case of Monitor, eclipses as well as transits
are of interest, but it is important to maximise the amount of
information that can be extracted from the light curve. All Monitor
campaigns were designed to ensure that the interval between
consecutive data points be less than 15\,min, and preferably closer
to 5\,min. The first value ensures that the duration of the shortest
occultations of interest ($\approx 1$\,h) is resolved, while the
second ensures that the \emph{ingress} and \emph{egress} is
resolved. The sampling rates are similar to those of other ground- and
space-based transit surveys, e.g.\ COROT.

Some of the telescopes of interest offer a queue scheduled service
program. This is not generally used for transit surveys because
there is no way to control the distribution of the observations in
time and because it is only possible to guarantee continuous observing
over a short duration -- generally 1\,h. The accepted wisdom has been
that one must observe continuously for at least the duration of an
occultation to ensure that events are observed completely enough. On
the other hand, service mode observations present a significant
advantage: they allow us to make use of the relatively lax observing
conditions requirements of our program. As relative rather than
absolute photometric accuracy is the key, and our fields are not
excessively crowded, the program can be carried out in moderate seeing
(up to 1.5\arcsec) and partial transparency. This makes it more
feasible to request large amounts of time on the appropriate
telescopes, which are generally heavily oversubscribed. It should also
improve the sensitivity to long periods, as the data will be spread
over an entire season, which is particularly relevant for secondary
science goals such as the search for photometric rotation
periods. Where such a mode was available, we therefore requested
queue-scheduled observations. 

There is a risk associated with such a decision, because any
occultation detected in this mode is likely to be incomplete, and the
number of distinct observations taken during a single occultation will
be small -- typically 4 at most. Additionally, observed occultations
may be separated by long periods without data, and our ability to
detect them by phase-folding the light curves will depend on the
long-term stability of the instrument and on the presence of any
additional long-timescale variability (e.g.\ starspots), which are
difficult to estimate a priori. Some of our target clusters will be
observed in both queue scheduled and visitor modes, and we will use
these datasets to perform an a posteriori evaluation of the relative
advantages of each mode for occultation surveys.

The total time allocation requested for each cluster was chosen to
ensure that the probability to observe at least three separate
occultation events for periods up to 5\,days should exceed 50\%. We
carried out Monte Carlo simulations of occultation observability under
various assumptions regarding the distribution of the observations in
time, including not only visitor mode observations with an adjustable
number of runs of variable duration, but also service mode
observations organised in fixed duration `blocks' distributed
semi-randomly. Interestingly, we found that, provided the time span of
the observations was significantly longer than the orbital periods
under consideration (we investigated periods up to 10\,d), the
relevant quantity was the total time $t_{\rm tot}$ spent on target,
and that a total of $\sim 100$\,h for each target was appropriate. We
therefore requested 100\,h per cluster in service mode. When only
visitor mode was available, we requested a number of nights totaling
up to slightly more than 100\,h to account for time lost to weather,
with the allocation being splits into two or more runs to ensure that
the time span of the light curves was significantly longer than 5
days.

The observations of each cluster are summarised in
Table~\ref{tab:obs}.  The telescope/instrument combinations used are:
the 2.2m MPI/ESO telescope (2p2) with the Wide Field Imager (WFI), the
2.4\,m Isaac Newton Telescope (INT) with the Wide-Field Camera (WFC),
the 3.6\,m Canada-France-Hawaii Telescope (CFHT) with the
MegaPrime/MegaCam camera, the 4\,m Blanco telescope at Cerro Tololo
Interamerican Observatory (CTIO) with the Mosaic~II imager, and its
northern twin the 4\,m Mayall telescope at Kitt Peak National
Observatory (KPNO) with the Mosaic imager.  The approximate magnitude
limits $I_{\rm sat}$ and $I_{\rm 5\%}$ and interval between
consecutive observations $\delta t$ were evaluated from the data
themselves wherever possible, and by analogy with other clusters
observed with the same set-up and strategy in the cases where data are
not yet available. The table clearly shows that, although the total
amount of time allocated to the survey in the vast majority of the
clusters matches or exceeds the requirement of 100\,h, the actual
amount of data collected often falls short of this requirement. This
is due to adverse weather conditions in the case of visitor mode
observations, and to lower completion rates than expected for the
queue-scheduled programs, due to the low priority assigned to
snapshot programs at the CFHT and technical problems delaying
Monitor observations on the ESO 2.2\,m. The sensitivity estimates
presented in Section~\ref{sec:sens} are re-computed after each
observing run (or delivery of data from service mode programs) and
used to evaluate whether an application for more data is needed.

\subsection{Data reduction and light curve production}
\label{sec:processing}

For a full description of our data reduction steps, the reader is
referred to Irwin et al.\ (in prep.). Briefly, we use the pipeline for
the INT wide-field survey \citep{il01} for 2-D instrumental signature
removal (crosstalk correction, bias correction, flat-fielding,
defringing) and astrometric and photometric calibration.  We then
generate the {\em master catalog} for each filter by stacking a few
tens of the frames taken in the best conditions (seeing, sky
brightness and transparency) and running the source detection software
on the stacked image. The resulting source positions are used to
perform aperture photometry on all of the time-series images. We fit a
2-D quadratic polynomial to the residuals in each frame (measured for
each object as the difference between its magnitude on the frame in
question and the median calculated across all frames) as a function of
position, for each of the detector CCDs separately. Subsequent removal
of this function accounts for effects such as varying differential
atmospheric extinction across each frame. We typically achieve a per
data point photometric precision of $\sim 2-5\ {\rm mmag}$ for the
brightest objects, with RMS scatter $< 1 \%$ over a dynamic range of
approximately 4 magnitudes in each cluster.

Photometric calibration of our data is carried out using regular
observations of \citet{lan92} equatorial standard star fields in the
usual way. This is not strictly necessary for the purely differential
part of a campaign such as ours, but the cost of the extra telescope
time for the standards observations is negligible, for the benefits of
providing well-calibrated photometry (e.g.\ for the production of
CMDs). In most of our target clusters we use CMDs for membership
selection, produced by stacking all observations in each of $V$ and
$i$ that were taken in good seeing and sky conditions (where possible,
in photometric conditions). The instrumental $V$ and $i$ or $I$
measurements are converted to the Johnson-Cousins system of
\citet{lan92} using colour equations derived from a large number of
standard star observations.

\subsection{Search for occultations}
\label{sec:analysis}

Although we intend to search for occultations in all of our
light curves in the long term, in the first instance we focus on
likely cluster members. In most cases, previous membership surveys --
based on proper motion, spectroscopy or photometry -- are not as deep
as our CMDs, and therefore we carry out our own membership selection.

In general, neither theoretical evolutionary models such as those of
\citet{bca+98} nor empirical sequences such as \citet{rg82} and
\citet{leg92} produce a good fit to the visible cluster sequence on
the $V$, $V-I$ CMD. Candidate cluster members are therefore selected
by defining an empirical main sequence, and moving this line
perpendicular to the mean gradient of the main sequence, toward the
faint, blue end of the diagram, by a constant adjusted by eye plus a
small multiple of the photometric error in $V-I$. The interested
reader is referred to \citet{iah+06} for more details and an example
of this procedure applied to our INT observations of M34.

Before searching for occultations, one must circumvent a major
obstacle: the intrinsic variability that affects the light curves of
almost all stars in the age range of the Monitor clusters at the level
of a few mmag to a few percent. When the major source of this
variability is the rotational modulation of starspots, it leads to
smooth variations that can be adequately modeled and subtracted before
the occultation search proceeds. We routinely search for this
modulation, using a sine-fitting procedure \citep[see e.g.][]{iah+06},
prior to starting the occultation search. When the sine-fitting
process leads to a detection, a starspot-model is fitted to the light
curve(s) at the detected period, following the approach of
\citet{dor87}. Even in the absence of a direct detection of rotational
modulation, any variability on time-scales significantly longer that
the expected duration of occultations (i.e. variability on timescales
of a day and longer) is filtered out using Fourier-domain filters
\citep{caf03, ai04, mpb+05}.

After pre-filtering, we search the filtered light curves for
occultations automatically using an algorithm based on least-squares
fitting of two trapezoid occultations of different depths but
identical internal and external durations, where the internal and
external durations are the time intervals between the $2^{\rm nd}$ and
$3^{\rm rd}$ contact, and the $1^{\rm st}$ and $4^{\rm th}$ contact,
respectively (Aigrain et al.\ in prep.). Note that a single box-shaped
transit is a special case of the double trapezoid where one of the
occultations has zero depth and the internal and external durations
are the same. The double trapezoid algorithm directly provides an
estimate of the basic parameters of the occultation which can be used,
together with an estimate of the primary mass based on its optical and
near-IR (2MASS) magnitudes if available, for a preliminary estimate of
the radius ratio of any detected systems.

In our youngest clusters, a variable fraction of the light curves are
affected by rapid, semi-regular variability which is not clearly
periodic (and hence not effectively detected by the sine-fitting
procedure or removed by the starspot fit) and overlaps with any
occultation signal in frequency space (and hence is not removed by the
Fourier domain filters unless at the expense of the signals of
interest themselves). We have found -- from our preliminary
investigation of the ONC data collected to date -- that visual
inspection of the light curves is the most effective way of finding
occultations in such cases. Those variations are thought to arise from
accretion- and time-varying activity-related effects, and concern
primarily classical T~Tau stars. The fraction of stars with inner
discs (estimated from near-IR excesses, \citealt{hll01}) therefore
provides a rough estimate of the fraction of light curves of members
we should expect to display such variability: $\sim60$\% for the ONC,
$\sim10$\% for NGC\,2362, and very small for all other clusters.

The calculations described in the present paper assume the double
trapezoid fitting algorithm is used in all cases, and essentially
ignore the effects of any variability that is not removed by the
pre-filtering stage. One should keep in mind that most of the occultations
we expect will be deep, and therefore their detectability will be
unaffected even by variability at the level of a few percent. The
events for which we do expect residual variability to play a
role are transits of planets around the youngest and highest mass
stars -- and we shall see in Section~\ref{sec:rvlim} that these are
also those for which the limitations imposed by the need to carry out
radial velocity follow-up are most stringent. We do plan to estimate
in detail the impact of residual intrinsic variability on our ability
to detect occultations by injecting artificial events into
the real pre-filtered light curves for each cluster, and repeating the
detection process, but this is beyond the scope of the present paper.

\section{Expected number of detections}
\label{sec:ndet}

A number of recent articles have explored the detection biases of
field transit surveys and the impact of the observing strategy on
their yield. \citet{pbm+05} analysed \emph{a posteriori} the planetary
transit candidates produced by the OGLE II Carina survey
\citep{usz+02,ups+03} in the light of the results of the radial
velocity follow-up, and highlighted the impact of correlated noise on
timescales similar to the duration of a typical
transit. \citet[][submitted]{pzq06} then developed a formalism to
model the correlated noise and account for its influence on expected
yields of transit surveys. In parallel, \citet{gau05,gsm05} developed
an analytical model of the number of detections expected for a given
survey. Both groups found the results of transit surveys to date to be
in agreement with those of radial velocity surveys, once the biases
were properly accounted for. \citet{pg05a} adapted the formalism of
\citet{gau00} to the special case of cluster transit searches, and
concluded that such searches may allow the discovery of `Hot Neptunes'
or `Hot Earths' from the ground \citep{pg05b}.

One might therefore envisage applying the prescription of
\citet{pg05a} to the Monitor project to evaluate the sensitivity of the
survey as a whole and in individual clusters. However, it relies on
certain assumptions which hold for planetary transits, but break down
for larger and more massive secondaries, as do all the analytical
methods proposed so far. For example, when dealing with stellar or
brown dwarf companions, the secondary mass is no longer negligible
compared to the primary mass. Also, grazing occultations -- usually
ignored in the planetary transit case as both rare and hard to detect
-- become common and detectable, and may in fact represent the
majority of detections. Another shortfall of published analytical
models is that they make simple assumptions regarding the noise
properties of the data, often assuming that the sole contributions to
the noise budget are source and sky photon noise, and that the noise
is purely white (uncorrelated). Similarly, the time array is generally
assumed to consist of regularly sampled `nights', making no allowance
for intermittent down time or time lost to weather. 

An alternative approach is to carry out Monte Carlo simulations once
the data have been fully reduced and analysed, injecting fake
occultation events into the light curves and applying the same
detection procedure as in the actual occultation search to derive
completeness estimates, and hence place constraints on the incidence
of companions in a given regime of parameter space. Using such an
approach, \citet{mss+05,wsb+05,bh06,bgd+06} derived the upper limits
on planet incidence in the field of a star cluster, although the
limits were not always stringent enough to constrain formation scenarios
because of insufficient target numbers and time sampling. Such a
procedure is envisaged in the long run for Monitor targets, but it
becomes very time consuming because of the huge variety of occultation
shapes and sizes which must be considered.

The purpose of the present work is to gain a global rather than
detailed insight into the potential of the Monitor project and the
type of systems which we may expect to detect. We have therefore opted
for an intermediate approach, based on a semi-analytical model, which
aims to incorporate important factors in the detection process, such
as the real time sampling and noise budget of the observations
(including correlated, or red, noise), while keeping complexity to a
minimum by carrying out calculations analytically wherever possible
and stopping short of actually injecting occultations into the observed
light curves. The calculations were implemented as a set of IDL
(Interactive Data Language) programs. Note that we have deliberately
excluded two significant factors, namely contamination by non-cluster
members and out-of-occultation variability, while the feasibility of
radial velocity follow-up is dealt with separately in
Section~\ref{sec:rvlim}.

\subsection{Procedure}
\label{sec:proc}

The number of detections expected in a given cluster is evaluated
according to:
\begin{equation}
N_{\rm det} = \int \int \int N_{\rm sys}(M) P_{\rm c} P_{\rm o} P_{\rm d}
\dd \log M \dd x \dd p
\end{equation}
where
\begin{itemize}
\item $M$ is the total system mass ($=M_1+M_2$ where $M_1$
and $M_2$ are the primary and secondary mass respectively);
\item $x$ is the parameter defining the companion. For binaries, we
  use the mass ratio $q=M_2/M_1$ as our defining parameter $x$,
  whereas for planet we use the companion radius $R_2$, for reasons
  explained below;
\item $p$ is the orbital period;
\item $N_{\rm sys}$ is the number of observed systems with total mass $M$;
\item $P_{\rm c}$ is the companion probability. Given that we
  parametrise our systems according to total mass rather than primary
  mass (for reasons which are explained below), this is the
  probability that a system of total mass $M$ is made up of at least 2
  components;
\item $P_{\rm o}$ is the occultation probability, i.e.\ the
  probability that eclipses or transits occur for a given binary or
  star-planet system;
\item  $P_{\rm d}$ is the detection probability, i.e.\ the
probability that occultations are detectable if they occur.
\end{itemize}

We use $M$ and $q$ to parametrise binaries, rather than
$M_1$ and $M_2$, for a number of reasons. Working in terms of $M_{\rm
  tot}$ is a more natural choice than $M_1$ because our surveys do not
resolve close binaries, and therefore each source detected on our
images must be considered as a potential multiple, rather than as a
single star. Additionally, most published mass functions for young
clusters (used for estimating $N_{\rm sys}$) are derived from CCD
surveys which do not resolve close binaries, and thus correspond to
systems rather than single stars. Using $q$ rather than $M_2$ is
convenient because the companion probability is often parametrised in
terms of mass ratio. Similarly, we use radius rather than mass to
parametrise the planets. In most cases, the mass of planetary
companions is generally negligible compared to that of their parent
stars, and the occultation and detection probabilities are thus
essentially a function of planet radius. Note that this is no longer
the case for the lowest mass primaries, so that we do not neglect the
planet mass in general. Instead, we assume a mass of $1\,M_{\rm Jup}$
for all planets in the simulations, whatever their radius, because the
mass-radius relation for young planets is both degenerate and ill
constrained, as discussed in Section~\ref{sec:mrrel_constraints}.

All quantities are computed over a three-dimensional grid of system
parameters. The total mass runs from $0.014\,M_{\sun}$ (i.e. just
above the planetary mass object limit) to $1.4\,M_{\sun}$ (the
approximate mass at which saturation occurs in the oldest, most
distant targets), in steps of $\dd\log M = 0.1\,M_{\sun}$.  The mass
ratio runs from 0 to 1 in steps of $\dd q = 0.05$. The planet radius
runs from $0.3\,R_{\rm Jup}$ to $2\,R_{\rm Jup}$ (most evolutionary
models of giant extra-solar planets assume initial radii in the range
2--$3\,R_{\rm Jup}$) in steps of $\dd\log R_2 = 0.1\,R_{\rm Jup}$. The
orbital period runs from 0.1 to 100 days for binaries, and 1 to 10
days for planets (no planets are currently known with periods less
than 1 day, while the occultation probability becomes negligible for
most planets with periods in excess of 10 days).

\subsection{Number of systems $N_{\rm sys}$}
\label{sec:nsys}

\begin{table*}
  \begin{minipage}{\linewidth}
    \caption{Mass range and number of cluster members monitored in each target cluster.}
    \label{tab:nobj}
    {\footnotesize{\bf (a):} Cases where $N$ was derived directly from the data.}
    \begin{center}
      \begin{tabular}[c]{@{}lrrrr}
        \hline
        Name & $\Omega$ & $M_{\rm L}$ & $M_{\rm H}$ & $N$ \\
        & ({\sq\degr}) & ($M_{\sun}$) & ($M_{\sun}$) & \\
        \hline
        NGC\,2362 & $0.38$ & $0.07$ & $1.14$ & 587 \\
        $h$ \& $\chi$\,Per (KPNO) & $0.60$ & $0.33$ & $1.49$ & 7756 \\
        NGC\,2547 & $0.56$ & $0.06$ & $0.88$ & 334 \\
        M50 & $0.38$ & $0.18$ & $0.88$ & 1942 \\
        NGC\,2516 & $1.13$ & $0.08$ & $0.56$ & 1214 \\
        M34 (INT) & $0.29$ & $0.16$ & $0.99$ & 414 \\
        \hline
      \end{tabular}
    \end{center}
    {\footnotesize{\bf (b):} Cases where $N$ was derived from the literature or from Monitor data taken with other telescopes.}
    \begin{center}
      \begin{tabular}[c]{@{}lrrrlrrrrrrr}
        \hline
        Name & $N'$ & $\Omega'$ & $M_{\rm L'}$ & $M_{\rm H'}$ & Ref & $\Omega$ & $f_{\Omega}/f'_{\Omega}$ & $M_{\rm 5\%}$ & $M_{\rm sat}$ & $N$ \\
        & &  ({\sq\degr}) & ($M_{\sun}$) & ($M_{\sun}$) & & ({\sq\degr}) & & ($M_{\sun}$) & ($M_{\sun}$) & \\
        \hline
        ONC & 500 & $0.07$ & $0.02$ & $0.5~$ & b & $0.28$ & $4.0$ & $0.04$ & $0.75$ & 2143 \\
        $h$ \& $\chi$\,Per (CFHT) & 7756 & $0.60$ & $0.33$ & $1.49$ & (KPNO) & $1.0$ & $1.0$ & $0.33$ & $1.49$ & 7756 \\
        IC\,4665 & 150 & $4.0~$ & $0.03$ & $0.2~$ & h & $4.0~$ & $1.0$ & $0.04$ & $0.45$ & 216 \\
        Blanco\,1 & 300 & $2.3~$ & $0.03$ & $0.6~$ & i & $1.17$ & $0.5$ & $0.06$ & $0.8~$ & 148 \\
        M34 (CFHT) & 414 & $0.29$ & $0.16$ & $0.99$ & (INT) & $1.0~$ & $1.9$ & $0.11$ & $0.7~$ & 845 \\
        M34 (KPNO) & 414 & $0.29$ & $0.16$ & $0.99$ & (INT) & $0.35$ & $1.3$ & $0.11$ & $0.7~$ & 560 \\
        \hline
      \end{tabular}
    \end{center}
    {\footnotesize{\bf Notes:} In the case of the ONC, we have taken
      the membership estimate from \protect\citet{hc00}, rather than
      \protect\citet{hil97} (which yields a smaller value), because
      the former corresponds to a mass range more similar to that of
      the Monitor targets, even though it covers only the central
      region of the cluster. If the discrepancy is arises from mass
      segregation (which would lead to a deviation from the log-normal
      MF adopted to compute $N$ \protect\citealt{hc00}), then the
      estimate we used should be close to the true number. It is also
      within 5\% of the number of detected sources in our field, which
      is reassuring given that few background or foreground field
      sources are expected in this case.}
  \end{minipage}
\end{table*}

Within the field of the observations, the expected number of observed
system of mass $M$ is
\begin{equation}
N_{\rm sys} (M) \, \dd\log M = N_{\rm C} \, f_{\Omega} \,  
\frac{\dd n}{\dd \log M} \, \dd \log M,
\end{equation}
where $N_{\rm C}$ is the total number of systems in the
cluster, $f_{\Omega}$ accounts for the fact that the field-of-view of
the observations may not cover the entire cluster, and
$\displaystyle\frac{\dd n}{\dd \log M}$ is the normalised
system mass function.

Assuming a $1/r$ profile for the density of cluster members,
\begin{equation}
\label{eq:fom}
f_{\Omega} = \left\{ \begin{array}{lcl}
  1 & \textrm{if} & \Omega \geq \Omega_{\rm C} \\
  \Omega / \Omega_{\rm C} & \textrm{if} & \Omega < \Omega_{\rm C}
    \end{array} \right. ,
\end{equation}
where $\Omega$ and $\Omega_{\rm C}$ are the solid angle covered by the
present survey and the total solid angle covered by the cluster,
respectively. This assumes that the survey area and the cluster
overlap to the greatest possible extent, and the value of $\Omega$
used in practice may differ from that implied by the actual surveyed
area to account for departures from this assumption imposed by
detector shape or other considerations (such as the need to avoid very
bright stars in the centre of some of the clusters, which would
otherwise be saturated and contaminate large areas of the detector).

In practice, $N_{\rm C}$ is rarely known directly. However, on can
generally find in the literature or, where data is already available,
determine from the Monitor data itself, an estimate of the number $N'$
of members in a given solid angle $\Omega'$ and mass range $M_{\rm
  L'}$ to $M_{\rm H'}$, obtained from an earlier membership
survey. Assuming that the vast majority of the systems are unresolved
-- an assumption which holds for all photographic and most CCD
surveys, with typical pixel sizes $> 0.1\arcsec$ -- we then have:
\begin{equation}
\label{eq:nprime}
N' = N_{\rm C} \, f_{\Omega'} \int_{M_{\rm L'}}^{M_{\rm H'}}
 \frac{\dd n}{\dd \log M} \, \dd \log M,
\end{equation}
where $f_{\Omega'}$ is the analog, for the earlier survey, of
$f_{\Omega}$ for the present one. Therefore, we can write
\begin{equation}
N_{\rm sys} (M) \, \dd\log M = A \, N' 
\frac{\dd n}{\dd \log M} \, \dd \log M 
\end{equation}
where $A$ is a normalisation accounting for the difference in spatial
coverage and mass range between the previous survey and the present
one:
\begin{equation}
A = \frac{f_{\Omega}}{f_{\Omega'}} \,  
\left[ \int_{M_{\rm L'}}^{M_{\rm H'}} (\dd n / \dd \log M) \, \dd \log M \right]^{-1}
\end{equation}

In some cases, $N'$ is available from spectroscopic surveys,
but generally we use contamination-corrected numbers of candidate
members identified on the basis of their position in colour-magnitude
diagrams. Mass segregation is not taken into account. 

Several recent determinations of the mass function (MF) of young open
clusters \citep{mbs+03,jnb+04,dbp+06} have concluded that a log-normal
distribution is a good fit over the entire mass range of interest to
us, from $\sim1\,M_{\sun}$ to the brown dwarf regime:
\begin{equation}
\label{eq:dndm}
\frac{\dd n}{\dd \log M} \propto \exp \left\{ 
            \frac{ - \left[ \log \left( M / M_0 \right) \right]^2}
                            {2 \, \sigma_M^2} \right\}
\end{equation}
where $M_0$ is the mean mass and $\sigma_M$ the standard deviation,
and the constant of proportionality is chosen to ensure the
distribution is normalised to 1. Typical values for $M_0$ and
$\sigma_M$ are $0.25\,M_{\sun}$ and $0.52\,M_{\sun}$ respectively
\citep{mbs+03}.  A log-normal with similar parameter also provides a
good fit to the mass functions derived from Monitor data in the cases
of the clusters analysed to date (see e.g.\ \citealt{iah+06} for M34
and Irwin et al.\ in prep.\ for NGC\,2516). As these mass functions
result from photometric CCD surveys with moderate spatial resolution,
within the orbital distance to which an occultation survey like
Monitor is sensitive, multiple systems are blended. This mass function
therefore corresponds to the number of systems, which is the quantity
of interest to us. There is no need to apply a correction for
binarity, which would be required only if our aim was to calculate the
distributions of masses for single stars and the individual components
of multiple systems, as discussed by \citet{cha03a}.

Table~\ref{tab:nobj} lists the overall number $N$ of likely cluster
members monitored with useful photometric precision in each cluster,
that is between the mass limits $M_{\rm 5\%}$ and $M_{\rm sat}$
corresponding to the magnitude limits of Table~\ref{tab:obs}. Wherever
possible, $N$ was derived from the Monitor data itself, following a
photometric membership selection procedure described in detail for the
case of M34 in \citet{iah+06}, and including an approximate field
contamination correction based on the Galactic models of the
Besan\c{c}on group \citep{rrd+03}. Note that in the case of clusters
lying close to the Galactic plane, and hence in crowded fields, we
applied a red as well as a blue membership cut, though the former was
designed to include the cluster binary sequence. $N$ was obtained by
summing the total number of candidate members after contamination
correction over the mass range $M_{\rm 5\%}$ to $M_{\rm sat}$. It is
worth noting the very rich nature of the twin clusters $h$ \& $\chi$\,
Per. In a study of the high-mass ($M\geq4\,M_{\sun}$ population,
\citet{shm02} already pointed out that these appear to be roughly 6 to 8
times as rich as the ONC, a finding which is consistent with our own
estimate.

Where data is not yet available or not yet analysed, $N$ was computed
from the literature or from Monitor surveys of the same cluster with
other telescopes according to:
\begin{equation}
N = A
\int_{M_{\rm 5\%}}^{M_{\rm sat}} \frac{\dd n}{\dd \log M} \, \dd \log M.
\end{equation}
We note that there are sometimes large discrepancies -- a factor 2 or
more -- between the values of $N$ we derive from our own data and
those extrapolated from earlier surveys in the literature. These will
be discussed in more detail in papers dealing with each cluster in
turn, but may arise both from the differences in mass range and
spatial coverage between previous surveys and Monitor, and from the
uncertainty on the level of contamination of the candidate members
lists by field stars. The discrepancies occur in both directions,
i.e.\ membership estimates based on our own data are sometimes below
and sometimes above the estimate extrapolated from the literature, and
no clear systematic trend emerged from the comparison of the two sets
of estimates. The differences highlight the need for spectroscopic
confirmation of our photometric membership selection wherever
feasible, and for the time being one should treat the values given in
Table~\ref{tab:nobj} with caution.

\subsection{Companion probability $P_{\rm c}$}
\label{sec:pcomp}

\subsubsection{Binaries}
\label{sec:pcomp_bin}

\begin{figure}
  \begin{center}
    \epsfig{file=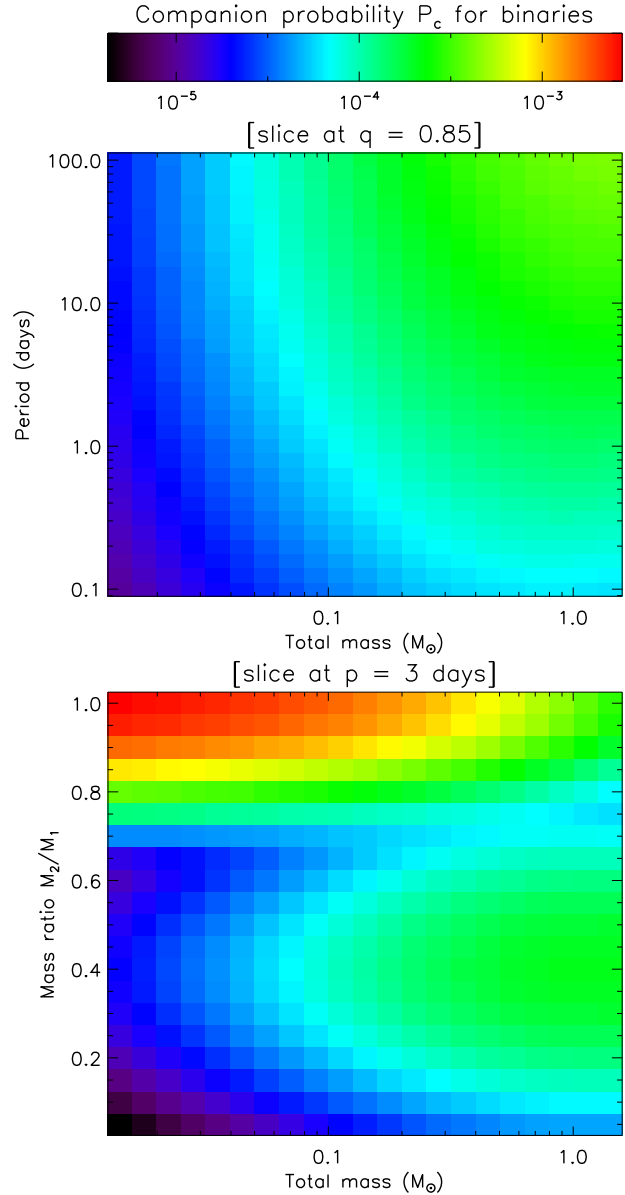,width=\linewidth}
  \end{center}
  \caption{Example two-dimensional cuts through the three-dimensional
    companion probability for binaries. Top: probability for mass
    ratio $q=0.85$ as a function of total mass and period. Bottom:
    probability for period $p=3$\,days as a function of total mass
    and mass ratio. The same logarithmic colour scale is used in both
    panels. \label{fig:pcomp_eb}}
\end{figure}

For stellar and sub-stellar companions, we use the following expression:
\begin{equation}
\label{eq:pceb}
P_{\rm c} (M, q, p) \, \dd q \, \dd p = 
f_{\rm c} \, \frac{\dd^2 p_{\rm c}}{\dd q \,\dd p} \, 
\dd q \, \dd p
\end{equation}
\noindent where $f_{\rm c}$ is the fraction of primaries of mass $M_1$
which host one or more stellar or sub-stellar companions, and $p_{\rm
  c}$ is the probability that a companion to such
a primary has mass ratio between $q$ and $q+\dd q$ and period between
$p$ and $p+\dd p$. $p_{\rm c}$ is normalised to 1 over
the range of $q$ and $p$ over which $f_{\rm c}$ is measured.

In a sample of 164 primaries of spectral type F, G or K, \citet{dm91}
found 62 binaries, 7 triples and 2 quadruples, corresponding to a
total companion fraction (over the entire period range, as both
spectroscopic and visual binaries were considered) of 43\%. These
authors also found that the distribution of orbital periods is well
fit by a log-normal function $\log(p_0)=4.8$ years and
$\sigma_p=2.3$\,d, noting an excess of short-period binaries ($1 \leq
P \leq 10$\,d) in a statistically young sample of Hyades
members. Combining field and open cluster samples of the same range of
spectral types, \citet{hmu+03} determined a companion fraction of 14\%
for orbital periods less than 10 years. This corresponds to a total
companion fraction of 47\% assuming the \citet{hmu+03} sample follow
the period distribution of \citealt{dm91}. \citet{hmu+03} also found
that the mass ratio distribution was generally bimodal, with a broad,
shallow peak stretching over $0.1 \la q < 0.75$, and a sharper peak
centred on $q \sim 1$, whose amplitude was about double that of the
other peak, and which was present for short-period binaries ($P <
50$\,d) only.

For primaries with masses below $0.5\,M_{\sun}$, the situation is less
clear. Imaging and radial velocity surveys of early M field dwarfs
(i.e. masses around $0.3$--$0.5\,M_{\sun}$) suggest that between 25\%
and 42\% have companions \citep{hm90,fm92,lhg+97,rg97}. For very
low-mass stars, the samples are smaller still. \citet{bbm+03,csf+03}
and \citet{scc+05} use high resolution imaging and find companions to
around 10-–20\% for objects with primary masses around $0.1\,M_{\rm
  sun}$; again for field dwarfs. However, spectroscopic samples
suggest that these studies may be missing significant numbers of close
in companions. \citet{mj05} examine a small sample of radial velocity
measurements, and estimate that accounting for systems with $a<3$\,AU
could increase the overall VLM star/BD binary frequency up to
32--45\%. \citet{br06} survey 53 VLM stars with Echelle spectroscopy,
and conclude that the overall binary fraction could be as high as
36\%. \citet{pjs05} argue that there is growing evidence that very low
mass binaries tend to have shorter periods than their higher mass
counterparts, and their mass ratio distribution is more strongly
peaked towards $q>0.75$.

Given the level of uncertainty on the multiplicity of low mass stars,
we have adopted an overall companion fraction $f_{\rm c}=50\%$,
independent of total system mass. To reflect the trend for low-mass
binaries to have shorter period, we have adopted a modified log-normal
period distribution, normalised over the full period range from 0\,d
to $\infty$, where the mean period scales with the primary mass:
\begin{equation}
\label{eq:binp}
\frac{\dd p_{\rm c}}{\dd \log p} = 0.5 \exp \left\{ 
  \frac{ - \left( \log p - \log p_0 - 0.5 \log M + \log M'_0 \right)^2}
       {2 \, \sigma_p^2} \right\}.
\end{equation}
\noindent where $p_0$ and $\sigma_p$ are taken from \citet{dm91}, and
$M'_0=1\,M_{\sun}$ should be close to the most frequent mass for the
stars in the sample of \citet{hmu+03}. For the mass ratio
distribution, we have adopted a double Gaussian, the lower peak's
amplitude also scaling with primary mass:
\begin{equation}
\label{eq:binq}
\frac{\dd p_{\rm c}}{\dd q} \propto  \left( \frac{M_1}{M_{\sun}}\right) 
     \exp \left[ \frac{ - \left( q - q_{0,1} \right)^2 }
                      { 2 \, \sigma_{q,1}^2} \right] +
     \exp \left[ \frac{ - \left( q - q_{0,2} \right)^2 }
                      { 2 \, \sigma_{q,2}^2} \right].
\end{equation}
\noindent where we have used $q_{0,1}=0.4$, $\sigma_{q,1}=0.2$,
$q_{0,2}=1.0$ and $\sigma_{q,2}=0.1$, which approximately reproduces
the distribution observed by \citet{hmu+03} in the case of F, G,
\& K primaries, and ensures that the amplitude of the second peak is
negligible for lower mass primaries. The constant of proportionality is
chosen to ensure normalisation over the mass ratio range 0--1. 

Two-dimensional cuts through the resulting three-dimensional companion
probability for binaries are shown in Figures~\ref{fig:pcomp_eb}. The
top panel illustrates the effect of the decreasing mean period towards
low masses, while the gradual disappearance of the low mass ratio peak
is visible in the bottom panel.

\subsubsection{Planets}
\label{sec:pcomp_pl}

The incidence and parameter distribution of planetary companions are
not yet well known, particularly around the low mass stars that
constitute the bulk of the Monitor targets. However, basic trends are
beginning to emerge from the results of radial velocity and transit
surveys to date, which concern primarily F, G and K stars. After
taking into account the period biases of both transit and radial
velocity searches, \citet{gsm05} found that the incidence of Hot
Jupiters (Jupiter-mass planets in 3 to 10 day orbits) around Sun-like
stars is roughly 1\%, while that of Very Hot Jupiters (Jupiter mass
planets in 1 to 3 day orbits) is roughly 5 to 10 times lower.

On the other hand, Jupiter-mass planets around M-dwarfs are rarer than
around Sun-like stars. For example, \citet{mbf+01} found only 1
Jupiter-mass companion in 3 years of surveying 150 M-stars, while
\citet{lba04} suggest that lower-mass planets might be more common
around low-mass stars, based on simulations in which initial
circumstellar disc mass scales with final star mass -- which may also
imply that M-star planets tend to have shorter orbital periods.

The prescription adopted here is an attempt to reflect the
above trends. We assume that
\begin{itemize}
\item 1\% of systems with $M \geq 0.5\,M_{\sun}$, and 0.5\% of systems
  with $M < 0.5\,M_{\sun}$ contain a Hot Jupiter (i.e.\ a planet with
  $R_2 > 0.7\,R_{\rm Jup}$ and $3\,{\rm d} \leq p \leq 10$\,d);
\item 0.2\% of all systems contains a very Hot Jupiter (i.e.\ a planet
  with $R_2 > 0.7\,R_{\rm Jup}$ and $0.4\,{\rm d} \leq p <
  3$\,d). Note that, in the present work, we have allowed the `very
  Hot' planet population to extend in period space down to 0.4 rather
  than the usual boundary of 1\,d. This was done in order to
  investigate the sensitivity to planets with extremely short periods,
  but one should keep in mind that no exoplanets with periods below
  1\,d have been reliably detected in radial velocity to date;
\item 3\% of all systems contain a Hot or very Hot Neptune (i.e. a
  planet with $R_2 > 0.7\,R_{\rm Jup}$ and $0.4\,{\rm d} \leq p \leq
  10$\,d).
\end{itemize}
These are very crude assumptions, but they allow an order of
magnitude estimate of the number of expected detections.

\subsection{Occultation probability $P_{\rm o}$}
\label{sec:pocc}

The occultation probability is:
\begin{equation}
\label{eq:pocc}
P_{\rm o} = \frac{R}{a} = 
R \left( \frac{2 \pi}{p} \right)^{2/3} \left( GM \right)^{-1/3} 
\end{equation}
where $R=R_1+R_2$ is the sum of the radii of both components (where
the radius of the stellar and substellar objects is deduced from the
mass-radius relation and that of the planets is varied between 0.3 and
$2\,R_{\rm Jp}$) and $a$ is the orbital distance, deduced from $M$ and
$p$ according to Kepler's $3^{\rm rd}$ law.

To compute $P_{\rm o}$, we need to know the radius of both primary and
secondary, and hence we need to introduce a mass-radius relation for
stars and brown dwarfs. The evolutionary models which we use for this
purpose also provide a mass-magnitude relation, which will be used in
computing $P_{\rm d}$. Both relations are described below.

\subsubsection{Mass-radius-magnitude relation for binaries}
\label{mrirel_adopted}

\begin{figure*}
  \begin{center}
    \epsfig{file=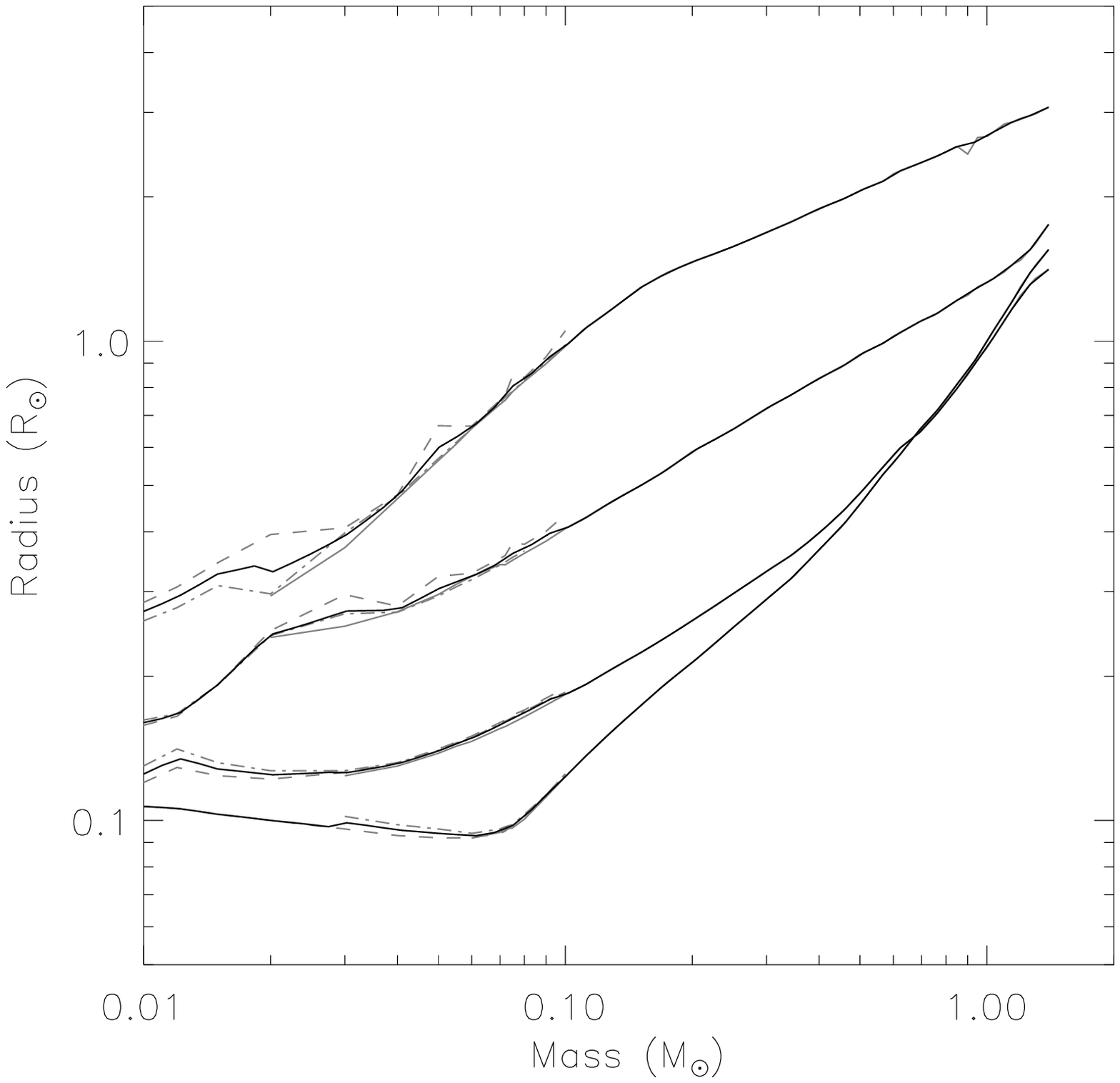,width=0.49\linewidth}
    \epsfig{file=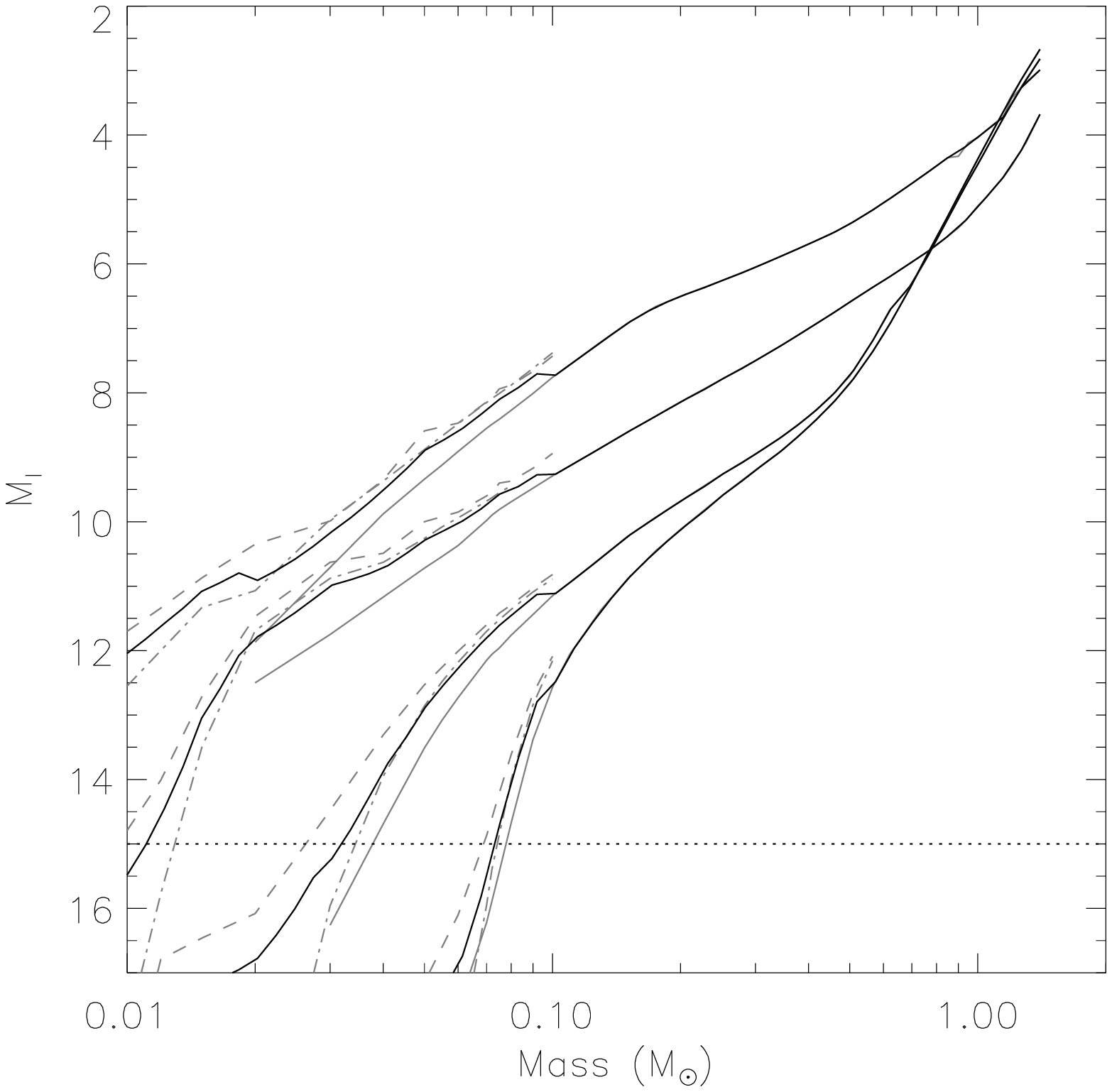,width=0.49\linewidth}
  \end{center}
  \caption{Mass-radius (left) and mass-absolute $I$-band magnitude
    (right) relations used in the calculations, shown here for 1, 10,
    100 and 1000\,Myr (from top to bottom). The model isochrones of
    \protect\citet{bca+98,cba+00} and \citet{bcb+03} are shown in grey
    (solid, dash-dot and dashed lines respectively) and the adopted
    relation in black. The horizontal dotted line in the right-hand
    panel shows the faint limit of our observations ($I\sim 19$) in
    the closest of our target clusters (Blanco\,1, $(M-m)_0 \sim
    7$). \label{fig:mrirel}}
\end{figure*}

To estimate the radius and absolute $I$-band magnitudes of stars and
brown dwarfs of a given mass, we interpolate over tabulated relations
between mass, radius and absolute magnitude which are illustrated for
selected ages in Figure~\ref{fig:mrirel}. The adopted relations at
each age were obtained by combining the NEXTGEN isochrones of
\citet{bca+98}, the DUSTY isochrones of \citet{cba+00}, and the COND
isochrones of \citet{bcb+03}, which span a total mass range
$0.5\,M_{\rm J}$ to $1.4\,M_{\sun}$, taking the mean of the values
predicted by the different models in the mass ranges of overlap. The
apparent $I$-band magnitude is then deduced from the absolute
magnitude using published values for the cluster distance $d$ and
reddening $E(B-V)$, using the extinction law of \citet{bm98}.

Although the DUSTY and COND models diverge strongly at low masses,
this occurs for very faint absolute magnitudes. As our deepest
observations reach no fainter than $I \sim 22$ and the closest target
cluster (Blanco\,1) has $(M-m)_0 \sim 7$, no detections are expected
for primaries with $M_I > 15$ (dotted line in
Figure~\ref{fig:mrirel}). Therefore, our very crude approach of taking
the average of the different models even where they diverge should not
strongly affect the results of the calculations.  However, one should
bear in mind that the mass-radius and mass-magnitude relations used in
the present work are indicative only, especially given the intrinsic
model uncertainties discussed in Sections~\ref{sec:mrrel_constraints}
and \ref{sec:young_binaries}.

\subsubsection{Behaviour of $P_{\rm o}$}
\label{sec:pocc_plots}

\begin{figure*}
\begin{minipage}{0.48\linewidth}
  \begin{center}
    \epsfig{file=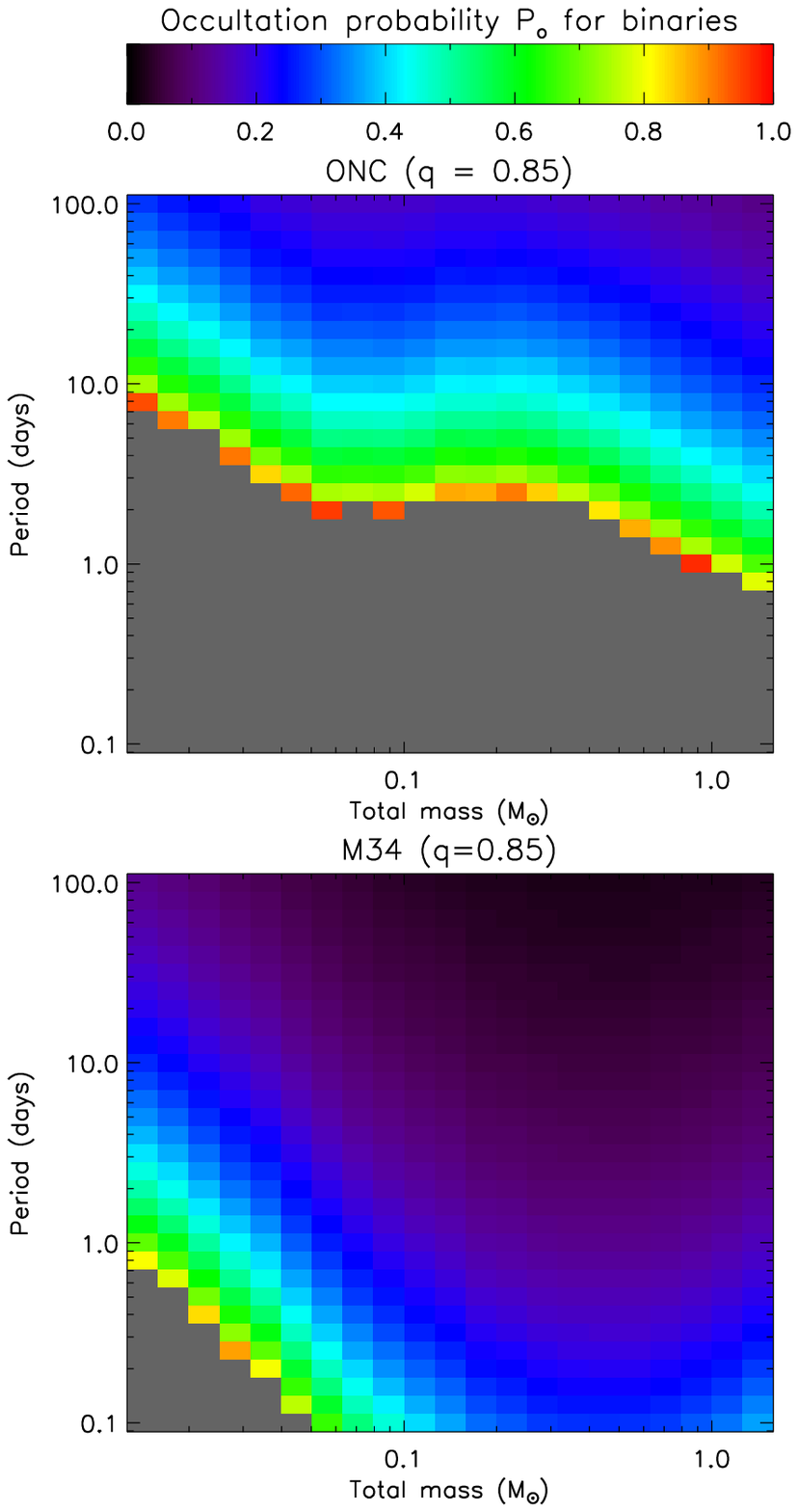,width=\linewidth}
  \end{center}
\end{minipage}\hfill
\begin{minipage}{0.48\linewidth}
  \begin{center}
    \epsfig{file=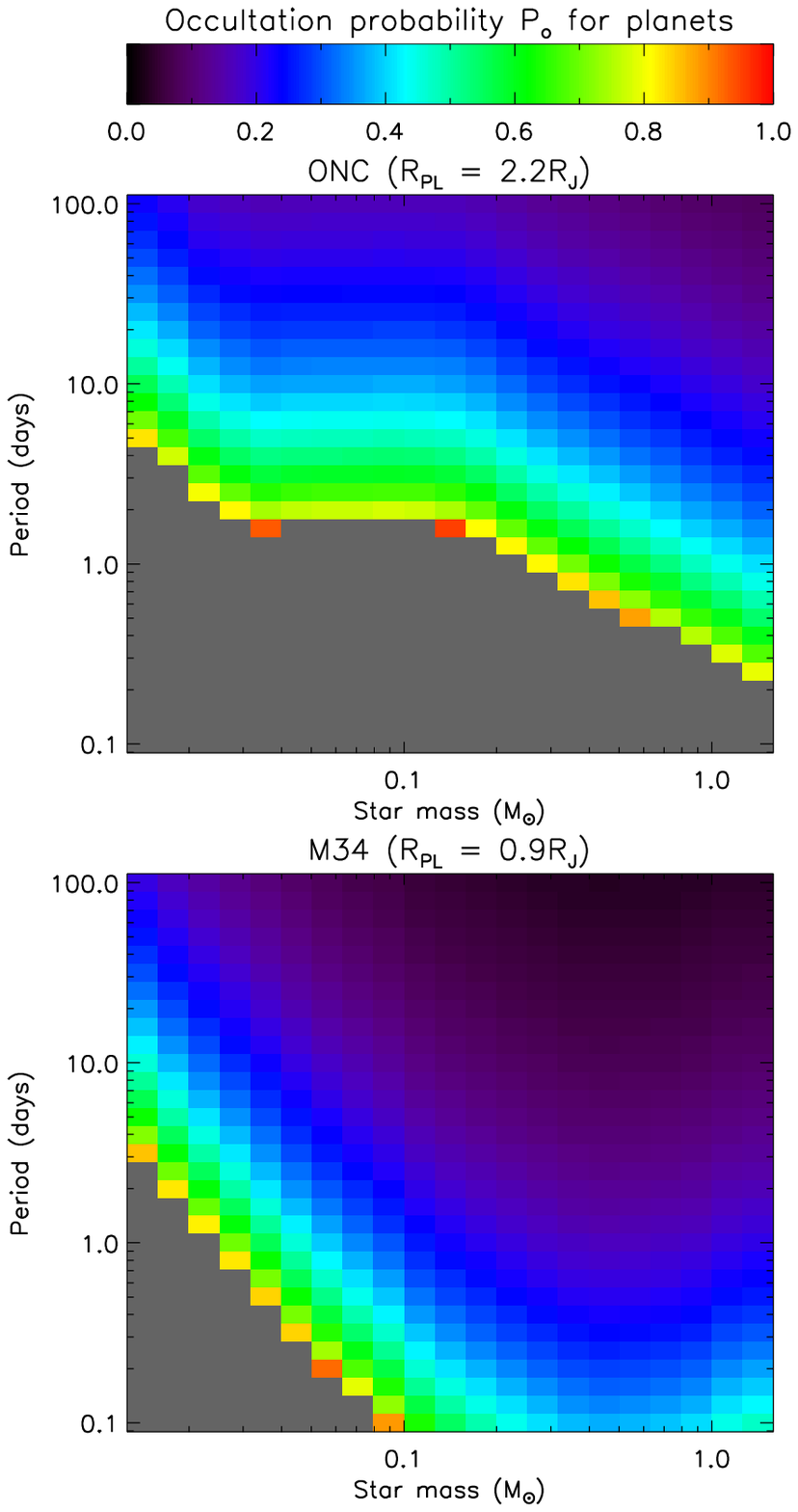,width=\linewidth}
  \end{center}
\end{minipage}
\caption{Example two-dimensional cuts through the three-dimensional
  occultation probability for the ONC (1\,Myr, top) and M34 (200\,Myr,
  bottom), for eclipses with mass ratio $q=0.85$ (left) and transits
  (right), for planet radii of $2.2\,R_{\rm Jup}$ (top) and
  $0.9\,R_{\rm Jup}$ (bottom). Regions shown in grey correspond to
  systems with orbital distance $a \leq R$. \label{fig:pecl}}
\end{figure*}

Having adopted a mass-radius relation, we are now in a position to
compute occultation probabilities for binaries, i.e.\ eclipse
probabilities. To do this, $M_1$ and $M_2$ are deduced from $M$ and
$q$, the mass-radius relation is used to give $R_1$ and $R_2$, which
are summed to give $R$. For planets, computing transit probabilities
involves deducing $M_1$ directly from $M$ (as all planets are assumed
to have mass $M_{\rm Jup}$), applying the mass-radius relation to give
$R_1$, and adding it to $R_2$ to give $R$. In both cases, $R$ is then
inserted back into Equation~(\ref{eq:pocc}) to give $P_{\rm o}$.
Figure~\ref{fig:pecl} shows 2--D cuts through the 3--D occultation
probability for two example clusters. Close inspection of this figure
highlights a few interesting points.
 
First, significant ($>0.2$)
occultation probabilities are encountered throughout much of the
parameter space of interest, even for planets (though this is no
longer true for planets with radii much below that of Jupiter). This
is due to the relative youth and low masses of the systems. As shown
on Figure~\ref{fig:mrirel}, young stars are larger than their main
sequence counterparts, while Kepler's $3^{\rm rd}$ law implies that
low-mass systems have smaller orbital distances -- for the same period
-- as their higher mass counterparts. Both of these effects tend to
increase occultation probabilities.

Equation~(\ref{eq:pocc}) also implies that, for a given total mass and
period, the occultation probability (and duration) \emph{increases}
towards lower companion masses throughout the range of companions for
which the mass-radius relation is relatively flat (up to $\sim
0.1\,M_{\sun}$ at 1\,Gyr). As the occultation depths will also be
comparable, this means that -- from the point of view of occultation
detection alone, and ignoring the incidence of such systems and the
constraints imposed by the need to perform RV follow-up -- we should
be at least as sensitive to transits as to eclipses in that regime.

Third, alignment considerations favour short period, low-mass
systems. These are particularly interesting because they offer very
stringent constraints for star formation theories. Finally, the
parameter space explored could nominally contain contact and
over-contact systems -- though the existence of such systems at such
early ages is far from established. This will need to be kept in mind
when modeling light curves in detail. For now, we set $P_{\rm o}=0$
for systems with $a \leq 2R$, to avoid counting these systems in the
overall detection rate estimates.

\subsection{Detection probability $P_{\rm d}$}
\label{sec:pdet}

The detection probability is evaluated using a Monte Carlo
approach. For each of $N_{\rm sim}$ realisations, we randomly select
an epoch and an inclination for the systems. Both the epoch and the
cosine of the inclination are drawn from a uniform distribution, the
latter being restricted to the the range $\cos i_{\rm min} \leq \cos i
\leq 1$ for which occultations occur, where $\cos i_{\rm min}$ is
directly deduced from $P_{\rm o}$:
\begin{equation}
\cos i_{\rm min} = \frac{R}{a} = P_{\rm o}
\end{equation}
For each realisation we evaluate whether the occultations would
have been detectable given the time sequence and noise
properties of the data. $P_{\rm d}$ is then the fraction of the
number of realisations in which the occultations would have been
detectable. 

\subsubsection{Detection statistic}

To evaluate the detectability of a given set of occultations, we
compute the detection statistic that it would give rise to, assuming
that a least-squares double-trapezoid fitting algorithm will be used
for detection. 

Among the most successful tools to search for occultations to date are
algorithms based on least-squares fitting of box-shaped transits
\citep{kzm02,ai04}. In such algorithms, the detection statistic to
maximise is often defined as the `signal-to-noise' $S$ of the transit,
which is the square-root of the difference in reduced chi-squared, 
$\Delta \chi^2$, between a constant model and the box-shaped transit
model:
\begin{equation}
S^2 = \Delta\chi^2 = 
\frac{\delta^2}{\sigma_{\rm w}^2 / n_{\rm t}}
\end{equation}
where $\delta$ is the transit depth, $\sigma_{\rm w}$ is the white
noise level per data point (assumed here to be the same for all data
points) and $n_{\rm t}$ is the number in-transit data points.

However, the true properties of the noise on occultation timescales
are generally not white. \citet{pzq06} have recently examined how
correlated noise on timescales of a few hours affects the
detectability of planetary transits, and proposed a modification of
this expression to account for correlated noise over a transit
timescale:
\begin{equation}
S^2 = \Delta\chi^2 = 
\frac{\delta^2}{\sigma_{\rm w}^2 / n_{\rm t} + \sigma_{\rm r}^2 / N_{\rm t}}
\end{equation}
where $\sigma_{\rm r}$ is the correlated noise over the
transit duration, and $N_{\rm t}$ is the number of distinct transits
sampled. 

Box-shaped transit-finding algorithms were designed with shallow
planetary transits in mind, whereas the majority of the occultations
expected in the context of the Monitor project will be deep and
grazing. A double-trapezoid occultation model, of which a single
box-shaped transit is a special case, will provide improved detection
performance (one can show that a box-shaped transit-search tool will
recover 94\% of the signal from a single triangular occultation,
\citealt{aig05}), and significantly better parameter estimation. Such
an algorithm will be described in more detail in Aigrain, Mazeh \&
Tamuz (in prep.). The detection statistic for such an algorithm is
defined as:
\begin{equation}
\label{eq:stat}
S^2 = \Delta\chi^2 =  
\frac{\delta_1^2}{\sigma_{\rm w}^2 / \Sigma_1 + \sigma_{\rm r}^2 / N_{{\rm t},1}} + 
\frac{\delta_2^2}{\sigma_{\rm w}^2 / \Sigma_2 + \sigma_{\rm r}^2 / N_{{\rm t},2}} 
\end{equation}
where $\delta_1$ ($\delta_2$) is the maximum depth of the primary
(secondary) occultation, $N_{{\rm t},1}$ ($N_{{\rm t},2}$) is the number
of distinct primary (secondary) occultations sampled, and $\Sigma_1$
($\Sigma_2$) is the sum of the weights attributed to the data points
in the primary (secondary) occultation. This sum is given by
\begin{equation}
\Sigma_x = n_{\rm c} + \sum \frac{2 \left( \tau_i - d_1 / 2 \right) }{d_2-d_1}
\end{equation}
where $n_{\rm c}$ is the number of points falling the central (flat)
part of the occultations, the summation runs over the remaining in-occultation
data points (which fall in the egress or egress) and $\tau_i$ is the
absolute deviation of the time of observation $i$ from the centre of
the occultation.

\citet{pzq06} find that a value of $S_{\rm lim} \sim 8$ is suitable for
typical survey parameters to define the level at which the false alarm
rate becomes unacceptably high. However, two additional requirements
were imposed. The first is that at least two distinct occultations be
sampled, which gives an upper limit to the orbital period. The second
is that a minimum of 4 in-occultation points be observed in total, so
as to provide a minimum of information on the shape of the event. Note
that \citet{af02} found that 4 in-transit bins provide the best
performance when using a step-function model with a variable number of
in-transit bins for detection purposes, which indicates that 4 samples
adequately describe the event (although in the present case, there is
no guarantee that the samples are evenly spaced within the
occultation).

\subsubsection{Occultation parameters}

In magnitude units, and if one ignores limb-darkening (recalling that
most Monitor data is obtained in $I$- or $i$-band where limb-darkening is
weak), occultations can be approximated as
trapezoids with a linear ingress, an optional flat-bottomed section,
and a linear egress. The occultation internal and external durations $d_1$
and $d_2$ are computed analytically assuming circular orbits
\citep[see e.g.][]{sm03}:
\begin{equation}
d_1 = \frac{p}{\pi} \, \arcsin \left[ \frac{1}{a \, \sin i}
    \sqrt{ \left( R_1 + R_2 \right)^2 - 
           \left( a \, \cos i \right)^2 } \right]
\end{equation}
\begin{equation}
d_2 =  \frac{p}{\pi} \, \arcsin \left[ \frac{1}{a \, \sin i}
    \sqrt{ \left( |R_1 - R_2| \right)^2 - 
           \left( a \, \cos i \right)^2 } \right].
\end{equation}

The occultation depths are evaluated numerically by constructing a
pixelised image of the disk of each component, with zero pixel values
outside the disk. The fraction of one disk hidden behind the other at
the centre of each occultation is evaluated by taking the pixel-by-pixel
product of the two images, appropriately positioned relative to each
other, and totaling up the number of non-zero pixels in the
result. The in-occultation flux is then obtained by subtracting from the
total out-of-occultation flux (the sum of disk-integrated fluxes from both
component) the fraction of the occulted star's flux that is hidden
from view. Again, limb-darkening is not taken into account.

\subsubsection{Noise level}
\label{sec:noise}

\begin{figure*}
  \centering
  \begin{center}
    \epsfig{file=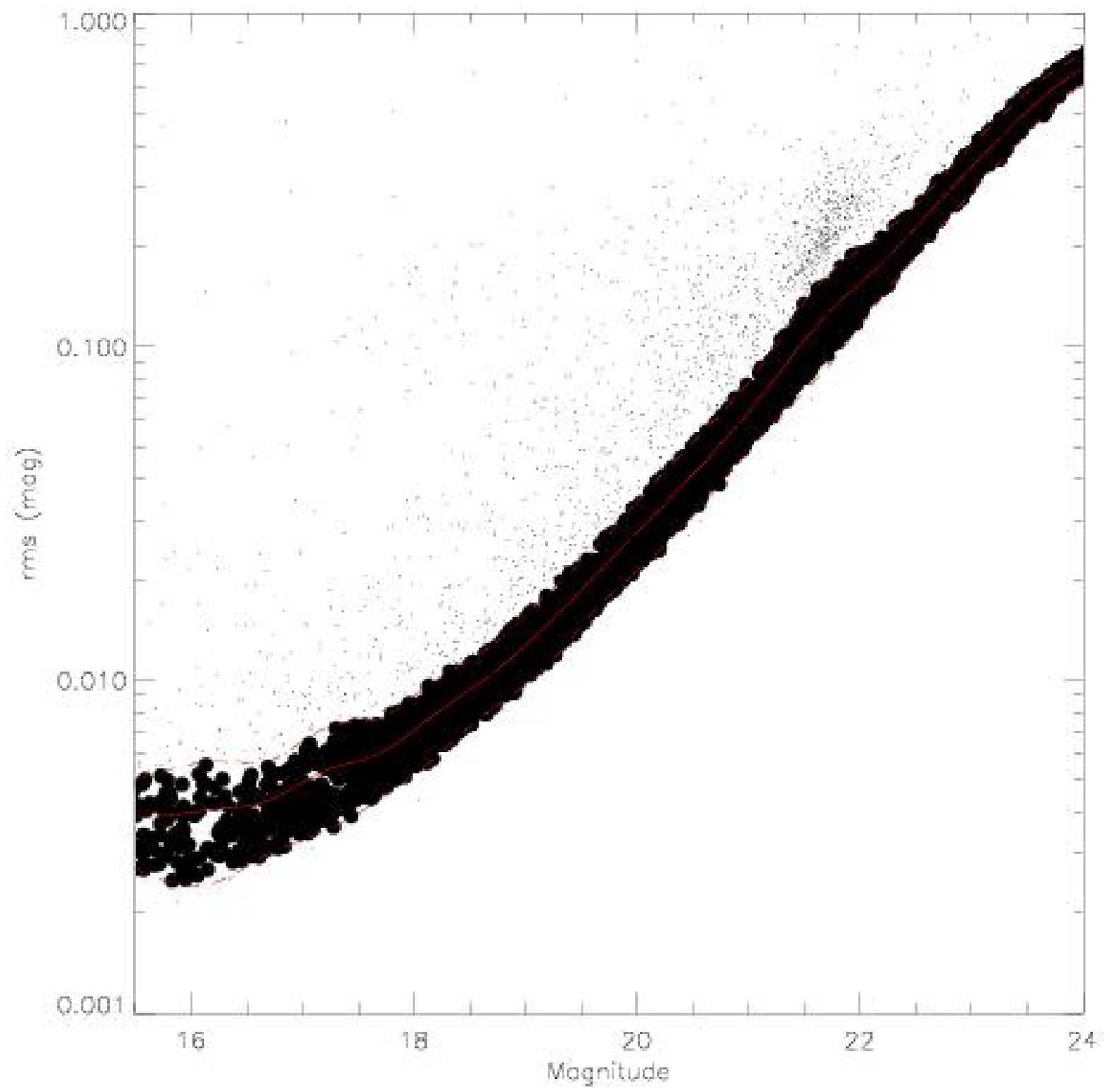,width=0.49\linewidth}
    \epsfig{file=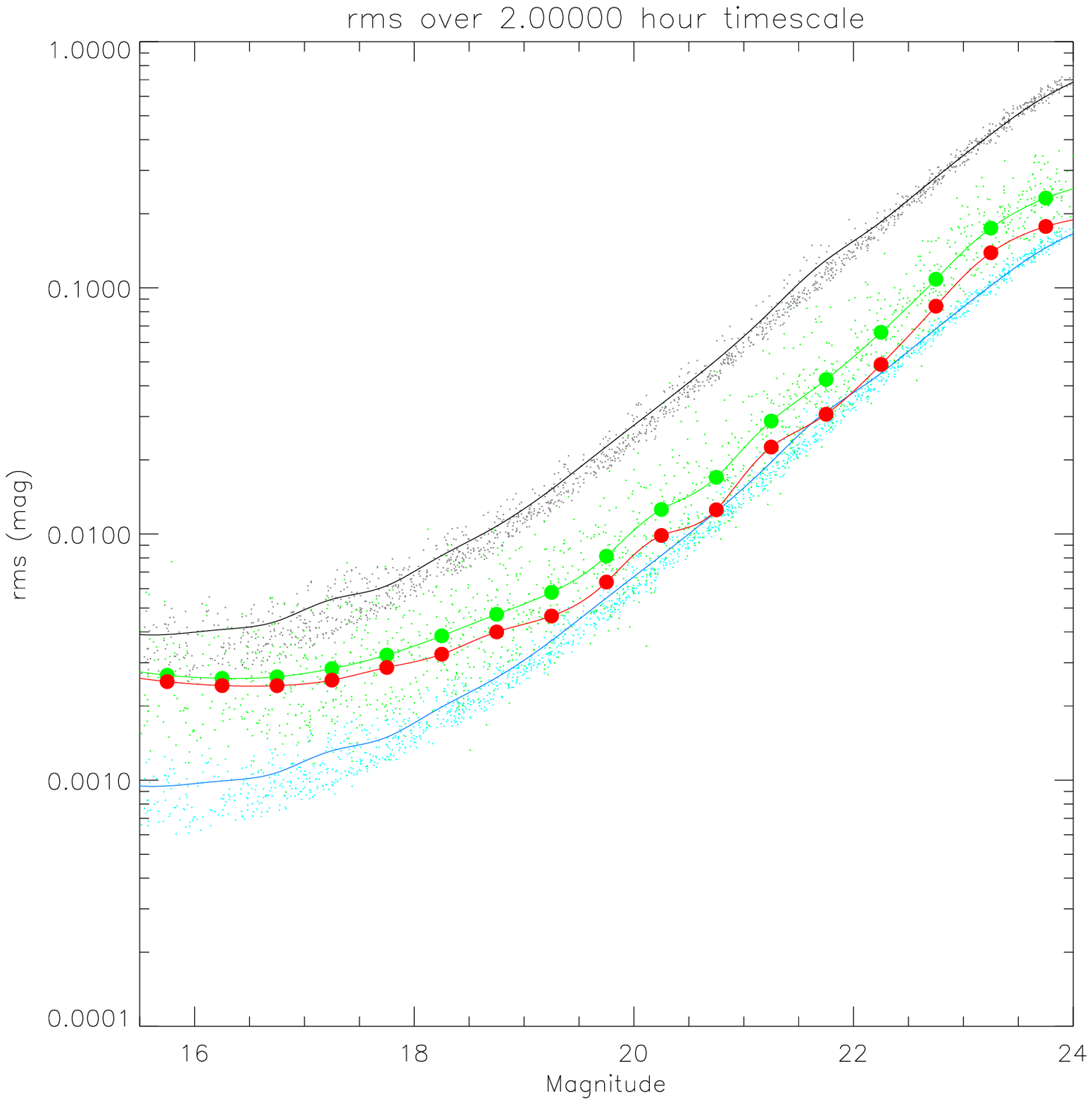,width=0.49\linewidth}
  \end{center}
  \caption{Left: Frame-to-frame rms $\sigma_{\rm w}$ versus magnitude
    for all M50 observations. Small black dots: all objects with
    stellar morphological classifications. Solid red line: spline fit
    to median rms versus magnitude. Dashed red lines: limits of the
    selection region for non-variable stars. Large black dots: objects
    selected as non-variable. Right: Noise budget over a 2\,h
    timescale. Small black dots: frame-to-frame rms for 100
    non-variable objects per 0.5 mag bin, selected at random. Black
    line: spline fit to median frame-to-frame rms versus
    magnitude. Blue dots and line: idem, divided by $\sqrt{n_{\rm
        int}}$. Green small dots, large dots and line: scatter
    $\sigma_{\rm s}$ of individual light curves over 2\,h times,
    median value in each magnitude bin, and fit thereto. Red small
    dots, large dots and line: idem for $\sigma_{\rm
      r}$. \label{fig:rms}}
\end{figure*}

We compute the noise budget of the light curves from our existing
data. The white and red noise levels $\sigma_{\rm w}$ and $\sigma_{\rm
  r}$ are both magnitude dependent, the latter also depending on the
occultation duration. Therefore, $\sigma_{\rm w}$ is evaluated once
per cluster, based on the median of the noise levels of the non
variable stars, and $\sigma_{\rm r}$ is evaluated once per cluster for
a range of likely occultation durations (from 0.5 to 3.5\,h)

First, we compute the median and scatter in each light
curve (using robust median-based estimators). We then sort the light
curves into 0.5\,mag wide bins of median magnitude, and compute the
median $\sigma_{\rm w}$ and scatter of the frame-to-frame rms in each
bin. A spline is then fitted to each quantity, and stars whose
$\sigma_{\rm w}$ fall within one sigma of the median rms fit are
selected as ``non variables''.

In each magnitude bin, we then select 100 non variable light curves at
random and use these to evaluate the average noise properties in that
bin. Each light curve is then smoothed over a number of timescales $d$
ranging from 30\,min to 3.5\,h, and we record the scatter $\sigma_{\rm
  s}$ of the smoothed light curve, counting only the intervals where
the smoothing window did not overlap with any data gaps. If a light
curve was affected by white noise only, we would expect $\sigma_{\rm
  s}=\sigma_{\rm w}\,n_{\rm int}^{-1/2}$, where $n_{\rm int}=d/\delta
t$ and $\delta t$ is the average interval between consecutive data
points. In general, $\sigma_{\rm s}$ is higher, owing to correlated,
or red, noise. For each bin, the average of the $\sigma_{\rm s}$ of
all 100 selected light curves is modeled as the quadrature sum of a
red and white noise components: $\sigma^2_{\rm r} = \sigma^2_{\rm s} -
\sigma^2_{\rm w} / n_{\rm int}$. The results of this procedure are
shown for M50 and $d=2$\,h in Figure~\ref{fig:rms}.

For each total system mass and, for binaries, mass ratio, the total
system magnitude is evaluated as follows. Applying the mass-magnitude
relation at the appropriate age to $M_1$ and $M_2$ and correcting for
the cluster distance and reddening, yields apparent $I$-band
magnitudes $I_1$ and $I_2$. The total system magnitude is then $I=-2.5
\log \left(10^{-0.4 I_1}+10^{-0.4 I_2} \right)$. If $I<I_{\rm sat}$ or
$I>I_{\rm sat}$, we set $P_{\rm d}=0$, to avoid counting saturated
systems or wasting time computing $P_{\rm d}$ for systems which are not
monitored sufficiently precisely to give a useful measurement of the
occultation depth. In all other cases, interpolating over spline fits
to the relations between magnitude and frame-to-frame and red noise
(over the most appropriate timescale, i.e. that closest to $d_1$)
yields the relevant values of $\sigma_{\rm w}$ and $\sigma_{\rm r}$,
which can then be inserted into into Equation~(\ref{eq:stat}).

\subsubsection{Behaviour of $P_{\rm d}$}
\label{sec:sens}

\begin{figure*} 
  \begin{center}
    \epsfig{file=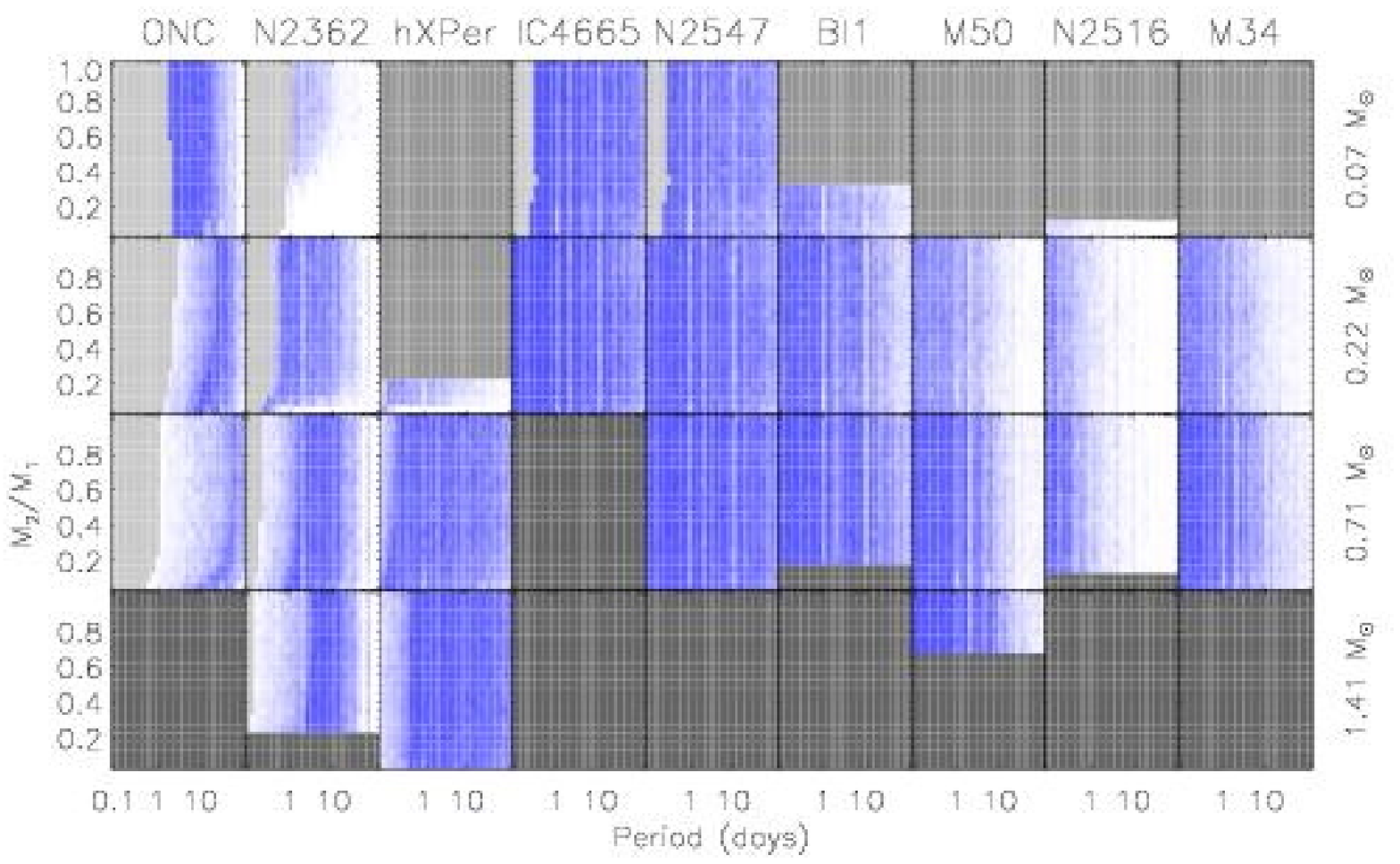,width=\linewidth}
    \epsfig{file=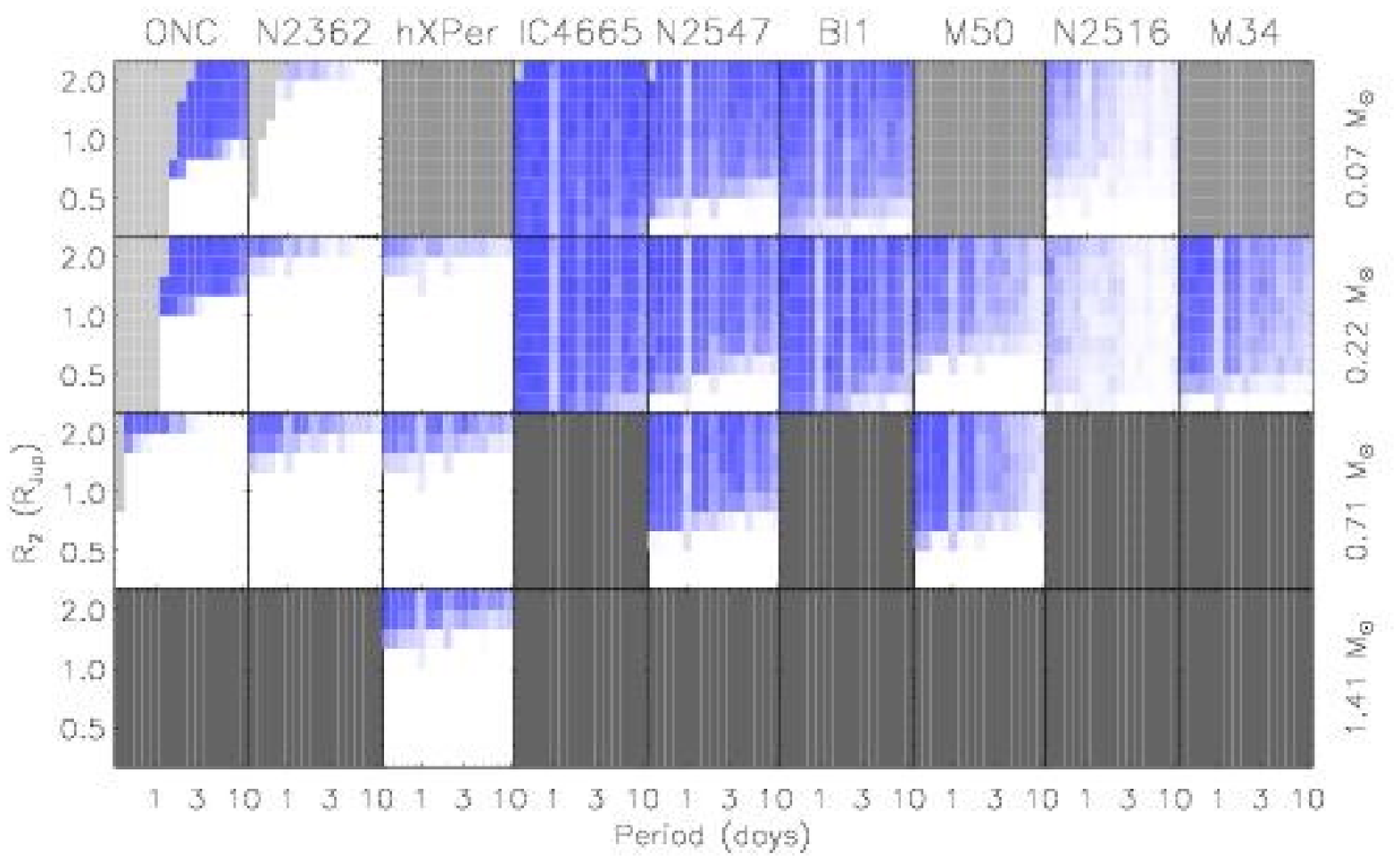,width=\linewidth}
  \end{center}
  \caption{Diagrams of $P_{\rm d}$ for binaries (top) and planets
    (bottom), as a function of orbital period (x-axis) and mass ratio
    or planet radius (y-axis) for each cluster (columns) and selected
    total system masses (rows). Blue areas correspond to detection
    probabilities close to 1 and the colour scale is linear. Areas
    shaded in light grey correspond to contact systems. Areas shaded
    in medium and dark grey correspond to systems that are too faint
    or saturated respectively. \label{fig:pdet}}
\end{figure*} 

The detection probabilities $P_{\rm d}$, computed as described
above, are illustrated for all the clusters in
Figure~\ref{fig:pdet}. They are shown as a function of period and
companion mass (eclipses) or radius (transits) for a variety of total
system masses.  These diagrams give a broad overview of the
sensitivity of our survey over the full parameter space.

At the time of writing, the photometric monitoring observations are
complete for three of our target clusters only. For all the other
cases, we used data from clusters observed in similar conditions to
estimate the noise properties, and added to or generated the time
sequence of observations artificially, ensuring the simulated time
sequence matches what is expected at least in the statistical
sense. Note that although we do have data from our INT campaign on the
ONC, we estimated the photometric precision based on observations of
M34 with the same telescope, because the intrinsic variability of ONC
stars makes it difficult to evaluate the true noise level from the ONC
light curves themselves. We used both the noise properties and the
time sequence of our CTIO campaign in NGC\,2516 to evaluate the
sensitivity of our KPNO survey in M34 and $h$ \& $\chi$\,Per, as the
telescopes and instruments are twins of each other and the observing
strategies closely matched. Wherever a given cluster was observed with
more than one telescope (m34, $h$ \& $\chi$\, Per), the simulations
were carried out separately for each
telescope. Figure~\ref{fig:pdet} then shows,
for each total mass, mass ratio or radius and period bin, the highest
of the sensitivities achieved with the different telescopes.

In each cluster, the survey was designed to ensure good sensitivity to
eclipses, and the sensitivity diagrams reflect this, with very good
sensitivity throughout much of the parameter space of interest for
binaries. Provided the period is short enough to accumulate the
minimum required number of observed in-transit points and transit
events, the eclipses of systems of all mass ratios are generally
easily detectable. It is only for the lowest total system masses that
mass ratio affects the sensitivity. For a given total mass, lower mass
ratio systems correspond to more massive (hence brighter) primaries,
counterbalancing the decrease in eclipse depth. When considering the
columns corresponding to $h$ \& $\chi$\,Per and M34, one should keep in
mind that the results shown are the combined results for several
surveys with different telescopes and observing strategies, which
leads to some discontinuities in the overall sensitivities.

In the clusters observed exclusively in visitor mode, we are sensitive
only to very short periods. As the eclipses are often deep and even a
single in-eclipse point can be highly significant, this short-period
bias is a consequence of the requirement that at least two separate
transit events be observed, rather than a direct detection limit. The
advantage of repeating observations after an interval of at least
several months is visible in the columns corresponding to the ONC,
NGC\,2362 and M50, where the sensitivity remains good up to $\sim 10$
days. The clusters observed in snapshot mode benefit from increased
sensitivity at long periods. However one should keep in mind that we
have not analysed data from any snapshot mode observations to
date. Snapshot mode observations may be affected by long-term
stability issues which will be hard to calibrate with the very patchy
time coverage we foresee. The performance of this mode compared to
more traditional visitor mode observations therefore remains to be
confirmed. In all cases, the complex period dependence of the
sensitivity is not fully resolved in the rather coarse grid we used,
but some of the ubiquitous sensitivity dips at exact multiples of 1
day, which are typical of transit surveys \citep{pbm+05,gsm05}, are
clearly visible. Because of the relatively coarse logarithmic period
sampling used, these dips are visible at 1, 2 and 10 day, where the
injected period was exactly a multiple of a day, but in reality they
would be present at all exact multiples of 1 day.

As expected, sensitivity to transits is much lower. The minimum
detectable planet radius is essentially a function of the cluster age
(which affects the stellar radii -- see ONC) and distance (which
affects the photometric precision -- see NGC\,2362 and
$h$ \& $\chi$\,Per). The importance of accumulating enough data is
highlighted by the very poor sensitivity in clusters observed for less
than the required 100\,h (NGC\,2516 and M34 at the high-mass end). It
is interesting to note that we are particularly sensitive to transits
around low-mass stars. We are limited to radii above that of Jupiter
in the youngest clusters, but this ties in with the expectation that
planets as well as stars are bloated at early ages. The sensitivity
peaks around M-stars, where planets with radii significantly below
that of Jupiter are detectable in some cases.

\subsection{Results}
\label{sec:ndet_res}

\begin{figure*}
  \centering
    \begin{center}
    \epsfig{file=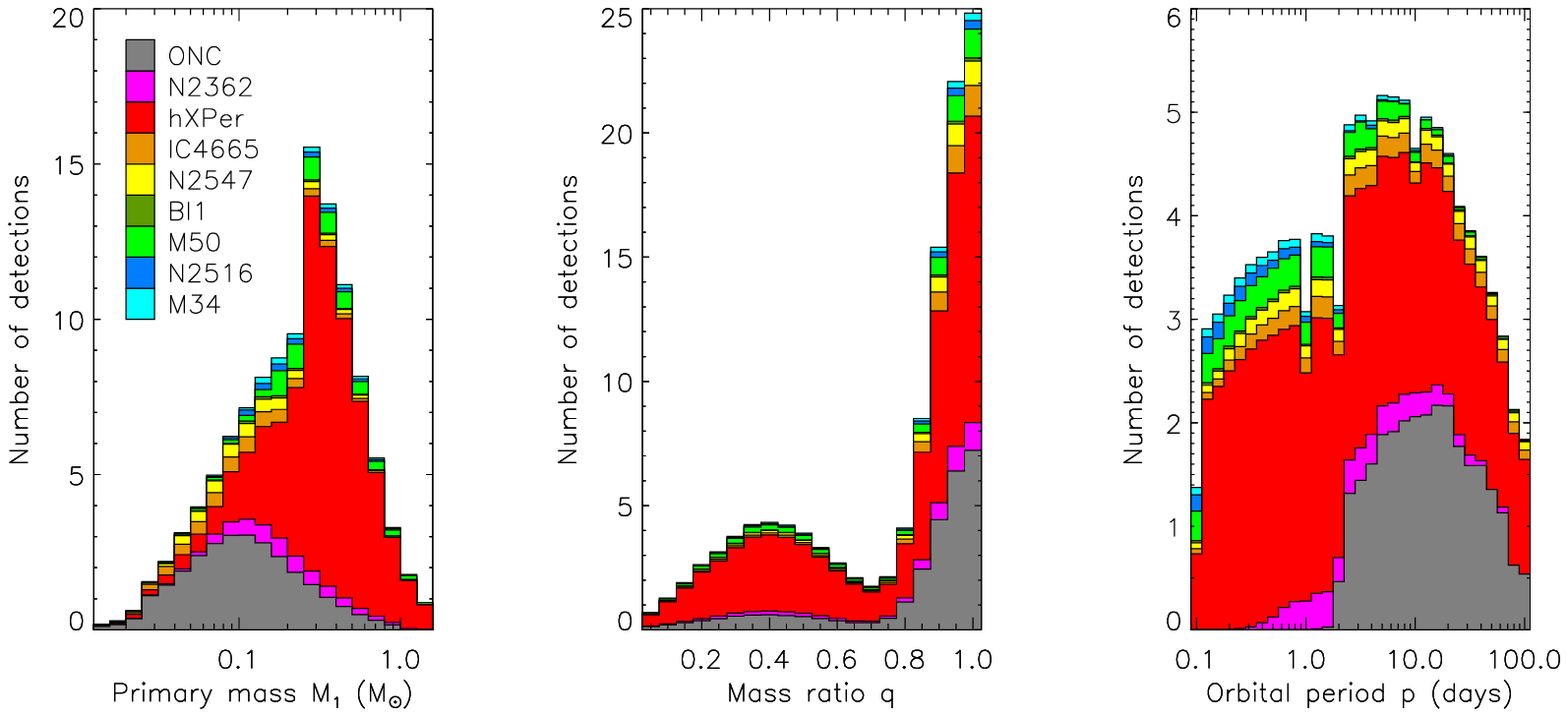,width=\linewidth}
    \epsfig{file=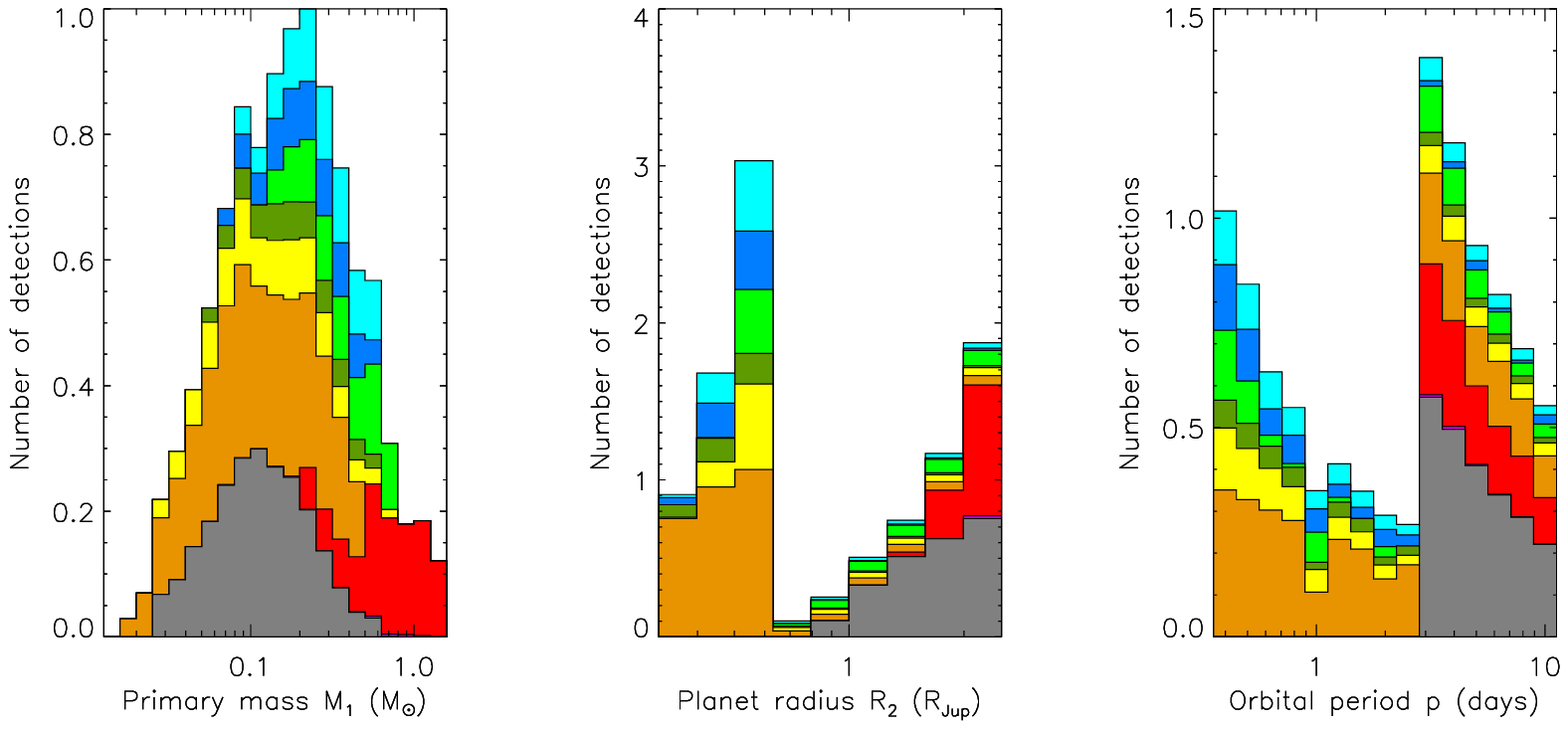,width=\linewidth}
  \end{center}
  \caption{Number of expected detections for binaries (top) as a function of primary mass (left), mass ratio (centre) and orbital period (right), and for planets (bottom) as a function of primary mass (left), planet radius (centre) and orbital period (right). The histograms are summed, each of the cluster being represented by a different colour, starting with the youngest (the ONC, in grey) at the bottom and ending with the oldest (M34, in light blue) at the top. The overall filled area corresponds to the total number of detections, all clusters combined.}
  \label{fig:ndet_hist}
\end{figure*}

\begin{table}
  \centering
  \begin{tabular}{lcccccc}
    \hline
    & \multicolumn{3}{c}{Binaries} & \multicolumn{3}{c}{Planets} \\
    Name & $N_{\rm c}$ & $N_{\rm o}$ & $N_{\rm d}$ & $N_{\rm c}$ & $N_{\rm o}$ & $N_{\rm d}$\\
    \hline
    ONC&   167.3&   57.3&   27.8&  135.0&   47.8&    2.3\\
    NGC\,2362&    45.6&   11.0&    4.7&   37.1&   11.3&    0.0\\
    $h$ \& $\chi$\,Per&   648.9&  106.0&   67.0&  631.9&  118.1&    1.2\\
    IC\,4665&    21.5&    5.8&    4.6&   14.3&    4.3&    3.1\\
    NGC\,2547&    30.4&    5.1&    3.9&   20.7&    3.9&    0.9\\
    Blanco\,1&    10.3&    0.9&    0.6&    8.0&    1.0&    0.5\\
    M50&   160.0&   12.7&    5.4&  127.0&   13.4&    0.8\\
    NGC\,2516&   103.0&    8.4&    1.5&   76.9&    8.4&    0.7\\
    M34&    45.1&    3.4&    1.4&   34.1&    3.5&    0.8\\
    \hline
    Total&  1230.7&  207.3&  114.0& 1084.4&  209.4&    8.6\\
    \hline
  \end{tabular}
  \caption{Expected number of binaries, eclipsing binaries and detectable eclipsing binaries, and of planets, transiting planets and detectable transiting planets, for each cluster and for the survey as a whole, under the assumptions described in the text.}
  \label{tab:ndet}
\end{table}

Table~\ref{tab:ndet} shows the number of observed cluster members
with companions, the number of observed occulting systems and the
number of detectable occulting systems for each cluster, for binaries
and planets separately. The former two are given by
\begin{equation}
\label{eq:nc}
N_{\rm c} = \int \int \int N_{\rm sys} P_{\rm c} ~ \dd \log M ~ \dd x ~ \dd P
\end{equation}
and 
\begin{equation}
\label{eq:no}
N_{\rm o} = \int \int \int N_{\rm sys} P_{\rm c} P_{\rm o} ~ 
\dd \log M ~ \dd x ~ \dd P
\end{equation}
where the integrals are performed on the entire range of parameter
space in the simulations, ignoring saturated or excessively faint
systems. The distribution of the detections in each cluster in terms
of primary mass, mass ratio (or planet radius) and orbital period is
shown in Figure~\ref{fig:ndet_hist}.

For the binaries, the main limiting factor is the number of systems
surveyed, which implies that the number of detections expected in some
clusters (e.g. Blanco\,1, NGC\,2516, M34) is of order unity, despite good
\emph{sensitivity} ($P_{\rm d}$ close to 1 over much of the parameter
space of interest). Given that each detection places a useful
constraint on a currently ill-constrained region of the mass-radius
relation, even those small numbers are interesting. On the other hand,
the number of expected detections in the rich twin clusters
$h$ \& $\chi$\,Per is very large, despite the fact that we are only
sensitive to companions to relatively massive stars. In the other
clusters, and \emph{a fortiori} for Monitor as a whole, significant
numbers of detections are expected. These should provide not only
strong constraints on the mass-radius relation over a range of masses,
but also imply that the Monitor survey will enable us to test
hypotheses regarding the binary fraction of low-mass stars at early
ages and the distributions of mass ratios and orbital periods for
young binaries. 

Figure~\ref{fig:ndet_hist} reflects the combination of the mass
function of each cluster, our assumptions about companion incidence
and the detection biases, mainly visible as a downward slope in the
period distribution. In the top row, corresponding to binaries, we see
two main types of behaviour. For the youngest clusters (the ONC and
NGC\,2362), we are sensitive primarily to low mass systems.  The mass
ratio distribution is consequentially dominated by the peak around
$q=1$, with very few low mass ratio systems. We predict few detections
with $p<1$\,d because these are contact systems at such early
ages. For the rest of the clusters, we have a more mixed picture,
including systems on both sides of the $M=0.5\,M_{\sun}$ boundary, and
our observations should thus enable us to test the hypotheses we
have made about the dependence of the mass ratio and period
distributions on total system mass (or primary mass).

We expect of order one planetary transit to be detected per
cluster. While this may seem like a low number, it is relatively high
compared to the amount of telescope time invested for a transit
survey, specially bearing in mind the particularly high potential
scientific impact of a transit detection in a young cluster.

As in the binary regime, there is also a clear distinction between the
youngest clusters (this time including $h$ \& $\chi$\,Per as well as
the ONC) and the older ones in the planet regime. As shown in the
bottom row of Figure~\ref{fig:ndet_hist}, in the ONC, we probe mainly
the Hot Jupiter population around low-mass stars, and we should be
able to test how much it differs from that around higher mass stars
and at later ages. The large size of the stars excluding both shorter
period systems (because of contact issues), and smaller planets
(because the transits are too shallow).  No detections are expected in
NGC\,2362 because it suffers from the same star-size issue as the ONC,
but the low-mass stars are too faint to allow us to detect Hot
Jupiters around them.  In $h$ \& $\chi$\,Per, the large number of
targets somewhat compensates the large cluster distance to give a
relatively high number of detections, and we are mainly sampling the
Hot Jupiter population around Sun-like stars, enabling a direct
comparison to the same population already well studied around older
field stars.  Together, the ONC and $h$ \& $\chi$\,Per constitute a
very populous (nearly 10\,000 targets) and interesting test-bed of
planet formation timescales, spanning as they do the full range of
circumstellar disk lifetimes. If no transits are detected in the ONC
or $h$ \& $\chi$\,Per, this will place a very strong upper limit on the
incidence of close-in giant planets at early ages.

On the other hand, for the older clusters, the majority of the
detections are expected in the very Hot Neptune regime. This is
because we have assumed, in a rather \emph{ad hoc} fashion, that very
Hot Neptunes are relatively common, whereas we have assumed, based on
observational evidence to date, that very Hot Jupiters are much
rarer. If very Hot Jupiters were more common than we have assumed, we
would detect them too. What Figure~\ref{fig:ndet_hist} really implies
is that, in the older clusters, we are almost exclusively sensitive to
the very Hot ($p<3$\,d) planet population, as are most other transit
surveys whether in the field or in clusters, but we are sensitive to
relatively small planets -- as noted by \citet{pg05b}. If very Hot
Neptunes are significantly rarer than we have assumed, we could easily
have no detections in any of this older group of clusters. Under the
set of assumptions used here, however, it is in IC\,4665 that the
largest number of transit detections of all the target clusters is
expected, despite the relatively low density of clsuter members,
because the host-star mass range we monitor are very favourable for a
transit survey. In this cluster, we are primarily sensitive to very
short period planets around very low-mass stars and brown dwarfs, and
this cluster will thus provide an interesting test of the abundance of
this type of planet.

In all clusters, the distribution of primary masses reflects mainly
the mass function of the cluster within the survey limits, and the
assumed difference in planet incidence between primaries above and
below $0.5\,_{\sun}$ is relatively hard to see except for
$h$ \& $\chi$\,Per. 

\section{Spectroscopic follow-up}
\label{sec:foll}

\subsection{Strategy}
\label{sec:foll_strat}

The light curve alone is not sufficient to ascertain the nature of any
companions detected through their occultations. Even if one assumes
that the primary lies on the cluster sequence, and that its mass and
radius are known, the light curve provides only an estimate of the
companion radius. Of course, photometrically selected candidates may
not in fact be cluster members. In addition, even if they are members,
the mass-radius-luminosity relations are so uncertain at early ages
and low masses that any photometric estimates of the primary mass and
radius could be highly unreliable. This last point is most valid for
the ONC, because of its youth, age spread, differential reddening and
the particularly low primary masses to which we are sensitive.

Multi-epoch spectroscopy is thus needed to ascertain the cluster
membership of any candidate systems and to determine the masses of the
components through radial velocity (RV) measurements. The relatively
faint nature of the target stars means that high-resolution
spectroscopy is extremely time consuming, and we have therefore opted
for a two-step follow-up strategy, consisting of at least one medium
resolution spectrum ($R \sim 5000$ to $10\,000$) on 2 to 4\,m class
telescopes, followed by multiple high resolution spectra ($R \sim
40\,000$) on 6 to 8\,m class telescopes for those candidates that
warrant it.

The number of RV epochs needed to constrain the companion mass is
minimised if a precise ephemeris for the occultations is
available. This requires several ($ \geq 3$) occultations to have been
observed, and -- unless the spectroscopic follow-up is carried out in
the same season as the original photometric survey -- occultations
observed in more than one season. For good candidates which do not
fulfill these criteria, we therefore foresee a photometric follow-up
stage, using telescopes with flexible scheduling and modest
field-of-view detectors, to attempt to observe additional
eclipses. This photometric follow-up can take place in the same time
frame as the first stage spectroscopic follow-up, as long as the
refined ephemeris is available when the RV data are analysed. These
observations are also used to attempt to detect secondary occultations
when the phase coverage of the initial observations did not allow
it. If carried out in multiple bandpasses, they can also be used to
refine the determination of the fundamental parameters of the
components.

Three main types of contaminants are foreseen:
\begin{itemize}
\item Background giants. In the case of most of our target clusters,
  any background giants would have to be outside the Galaxy to pass
  our membership cut. This is therefore not a major source of
  contamination, though medium-resolution spectroscopy allows the
  identification for most late-type giants through the measurement of
  gravity sensitivity features;
\item Background field dwarfs reddened onto the cluster
  sequence. These constitute the main source of contamination, but can
  be weeded out by comparing their spectral type to that expected from
  their optical and near-IR colours, because they do not follow the
  same colour-spectral type relation as cluster members;
\item Unreddened field stars which do follow a similar colour-spectral
  type relation to cluster members, but which lie in range of apparent
  magnitudes which allows them to pass our membership cut. This
  implies that they must lie in a rather restricted volume, mostly on
  the near-side of the cluster, and the number of contaminants of this
  type is not expected to be very large. A lack of youth indicators
  such as Lithium absorption lines and H$\alpha$ emission in the
  spectrum will be the main way of identifying these objects, together
  with dynamical indicators (systemic radial velocity incompatible
  with that of the cluster).
\end{itemize}
A single medium resolution spectrum in the red part of the visible
with a signal to noise ratio better than 20 per resolution element
typically requires less than 1\,h of exposure for a target with
$I<18.5$ on a 4\,m telescope, and yields spectral classification to
better than one subclass (based on \citet{khm+91} relative flux
indices). This is complemented by the search for youth indicators
(H$\alpha$ emission, Lithium absorption), though these are
expected to be present in some of our targets only. Gravity sensitive
lines and indices also provide some degree of discrimination between
young cluster members and giants or old field dwarfs.

This initial spectrum also typically provides a first epoch radial
velocity measurement at the few km/s level, which should suffice to
detect the variations induced by any stellar and most brown dwarf
companions (a $0.03\,M\,_{\sun}$ BD in a $10$\,d orbit around a solar
mass star will induce an RV amplitude of $3$\,km/s in the primary, and
shorter periods or larger mass ratio lead to increased
amplitudes). Unless the object has clearly been identified as a
non-member from the first spectrum, a small number of additional
medium resolution spectra are taken. If no RV variations are detected
at the km/s level, the object remains a good candidate provided the
depth of the occultations is consistent with a very low-mass companion
(otherwise, one must question the true nature of the occultations). If
variations are detected, they can be used to measure -- or at least
place a lower limit on -- the RV amplitude of the primary, and thus
the mass ratio of the system. Whether variations are detected or not,
this first set of measurements also provides an estimate of the
systemic radial velocity. Comparison of this with the cluster radial
velocity provides an independent test of cluster membership.

High resolution spectroscopy is then needed for all candidates that
survive the previous stage, i.e.\ those for which we derive a spectral
type and systemic RV consistent with cluster membership, and where
either we detect RV modulations, or the non-detection of RV
modulations is consistent with the minimum companion mass implied by
the light curve and our estimate of the primary mass, given the RV
precision achieved with medium resolution spectra.  In cases where we
have detected RV modulations, the goal of this second stage is to
resolve the secondary set of lines, and to obtain a full orbital
solution. In this case, the observations are best carried out in the
near-infrared (with instruments such as Phoenix on Gemini or CRIRES at
the VLT), where the contrast between primary and secondary is lower
than in the visible.

In cases where no RV modulation was detected so far, the second set of
lines is unlikely to be detectable, but increased RV precision is
needed to resolve the very low amplitude modulations of the
primary. For those systems, estimates of the mass and radius of the
primary must rely on relations between effective temperature (deduced
from the spectrum), mass and radius which, as we have seen in
Section~\ref{sec:intro}, are very poorly calibrated at early ages. If
high mass ratio systems are detected in the same cluster as low mass
ratio systems, the constraints the former will provide on these
relations will be used to refine the estimates of the parameters of
the primaries of the latter.

In the Section~\ref{sec:rvlim}, we investigate the expected RV
precision as a function of magnitude and rotational velocity with
medium resolution instruments on 4m class telescopes (such as EMMI on
the NTT or ISIS on the WHT) and with higher resolution instruments on
8m class telescopes (such as FLAMES and UVES on the VLT).

\subsection{Limits imposed by radial velocity accuracy}
\label{sec:rvlim}

Although the flux from both primary and secondary is maximised in the
near IR for low mass stars, precision RV work from near-IR spectra is
a relatively un-tested area. On the other hand, recent surveys in the
optical have generated a wealth of information on the achievable RV
performance. An additional advantage of the optical is the
availability of high resolution spectrographs with wide-field
multiplexing capabilities. Therefore, we investigate here the optical
only.

State of the art RV instruments are today reaching accuracies of a few
m/s, and are capable of detecting the modulations imparted on their
parent stars even by Neptune-mass planets \citep[see
e.g.][]{lmp+06}. However, this requires multiple, very high resolution
spectra to be obtained with relatively high signal-to-noise ratio, and
is hence feasible only for bright stars. Candidates from transit
surveys carried out on telescopes with apertures of 1\,m and above
tend to be much fainter, which makes their follow-up much more
difficult. However, the recent campaigns to follow-up OGLE candidates
\citep{bpm+05,pbm+05} have shown that it is possible to obtain reach
RV accuracies down to $\sim 100$\,m/s down to $I=17$ for non-rotating
stars with FLAMES on the VLT, using high resolution spectra
($R~20\,000$--$40\,000$) covering a broad wavelength range centred on
$\sim 600$\,nm, with a simultaneous ThAr reference for wavelength
calibration.

Many of the candidates expected in the context of Monitor are both
fainter and much redder than OGLE targets. Using a redder part of the
spectrum allows useful signal-to-noise ratios to be accumulated in
much shorter exposure times, at the cost of a smaller number of lines
and the loss of the simultaneous wavelength reference. In particular,
the Ca~{\sc ii} triplet around 850\,nm (hereafter Ca~T) is routinely used
for stellar population kinematic studies, yielding accuracies down to
2\,km/s in reasonable exposure times down to $I \sim 18$ using FORS on
the VLT \citep[see e.g.][]{pzg+04}. 

We have therefore computed the limiting RV accuracy achievable in 1\,h
exposure times as a function of magnitude and projected rotation rate
$v \sin i$ (rotation broadens the lines and limits the achievable
accuracy for three types of observations): 
\begin{itemize}
\item Ca~T observations using medium resolution
  spectrographs on 4\,m class telescopes, such as EMMI on the NTT or
  ISIS on the WHT (hereafter M850); 
\item Ca~T observations using higher resolution spectrographs on
  8\,m class telescopes, such as FLAMES on the VLT (hereafter H850);
\item observations in the 600\,nm region using higher resolution
  spectrographs on 8\,m class telescopes, such as FLAMES on the VLT
  (hereafter H600).
\end{itemize}
The M850 calculations enable us to check what fraction of the objects
for which we expect to detect eclipses we will detect the RV
modulations for in the first (medium resolution) stage of our
follow-up strategy (see Section~\ref{sec:foll_strat}). The detail of
the calculations and settings used for each type of observations is
given in Appendix~\ref{app:rvprec}.

\begin{figure*}
  \centering
  \epsfig{file=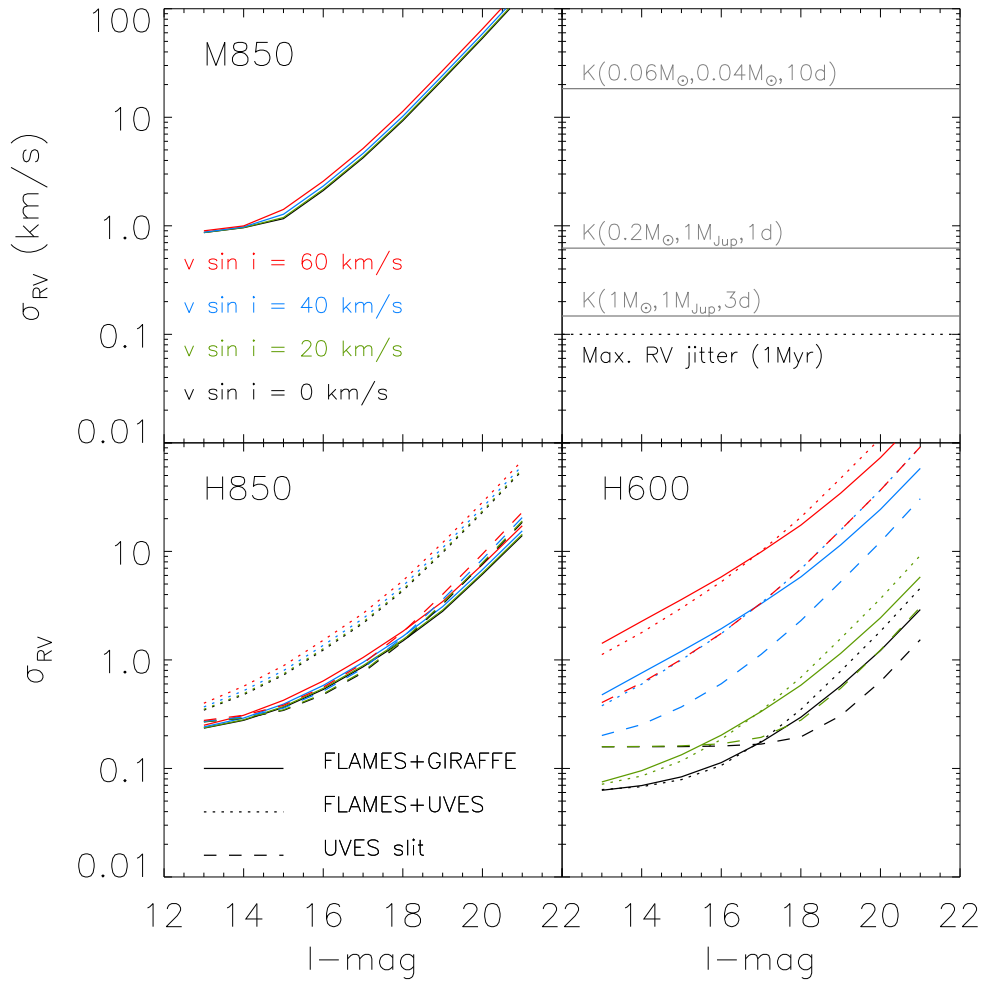,width=0.8\linewidth}
  \caption{Theoretical radial velocity error as a function of apparent
    $I$-band magnitude based on M850, H850 and H600 observations (top
    left, bottom left and bottom right respectively). All calculations
    are based on integration times of up to 1\,h taken in good
    atmospheric conditions (see text). The black, green, blue and red
    curves in each panel correspond to objects with $v \sin i$'s of 0,
    20, 40 and 60\,km/s respectively. In the top right panel, RV
    semi-amplitudes are shown as horizontal grey lines for the
    primaries of a number of example systems, labeled with the
    corresponding primary and secondary mass and orbital period. In
    the same panel, the black horizontal dotted line shows the
    activity induced RV jitter expected in the worst case (youngest,
    most active stars). \label{fig:rvprec}}
\end{figure*}

Figure~\ref{fig:rvprec} shows the results of this exercise for $v \sin
i$ ranging from 0 to 60\,km/s (i.e. rotation periods down to 0.8\,days
for a radius of $1\,R_{\sun}$). As expected, precisions of a few km/s
can be achieved using M850 observations down to $I\sim 17$. The
intrinsic width of the Ca~T lines makes them very insensitive to
rotation. To reach fainter objects or achieve better precision
requires larger telescopes and higher resolution instruments. With
those, still using the Ca~T, i.e.\ H850, it is possible to reach
precisions of $\sim2$\,km/s at $I\sim18$ and $\sim200$\,m/s at
$I\sim13$, while H600 observations can provide increased precision --
down to $\sim 50$\,m/s at the bright end -- provided the rotational
velocity is moderate ($\leq 30$~\,km/s). These calculations assume an
M2V spectral type and insignificant reddening.

This leads to the somewhat puzzling conclusion that neither spectral
region is globally optimal, and that the optimal strategy will have to
be selected on the basis of the spectral type and rotation periods or
$v \sin i$ of the individual candidates (where those are not measured,
rough estimates can be inferred from the candidate's age and spectral
type). The trade-off between the higher throughput of UVES in
slit-mode and the lower seeing-induced errors of FLAMES will depend on
the surface density of candidates in a given cluster. There appears to
be little gain at any magnitude in using FLAMES$+$UVES over
FLAMES$+$GIRAFFE.

\begin{figure*} 
  \begin{center}
    \epsfig{file=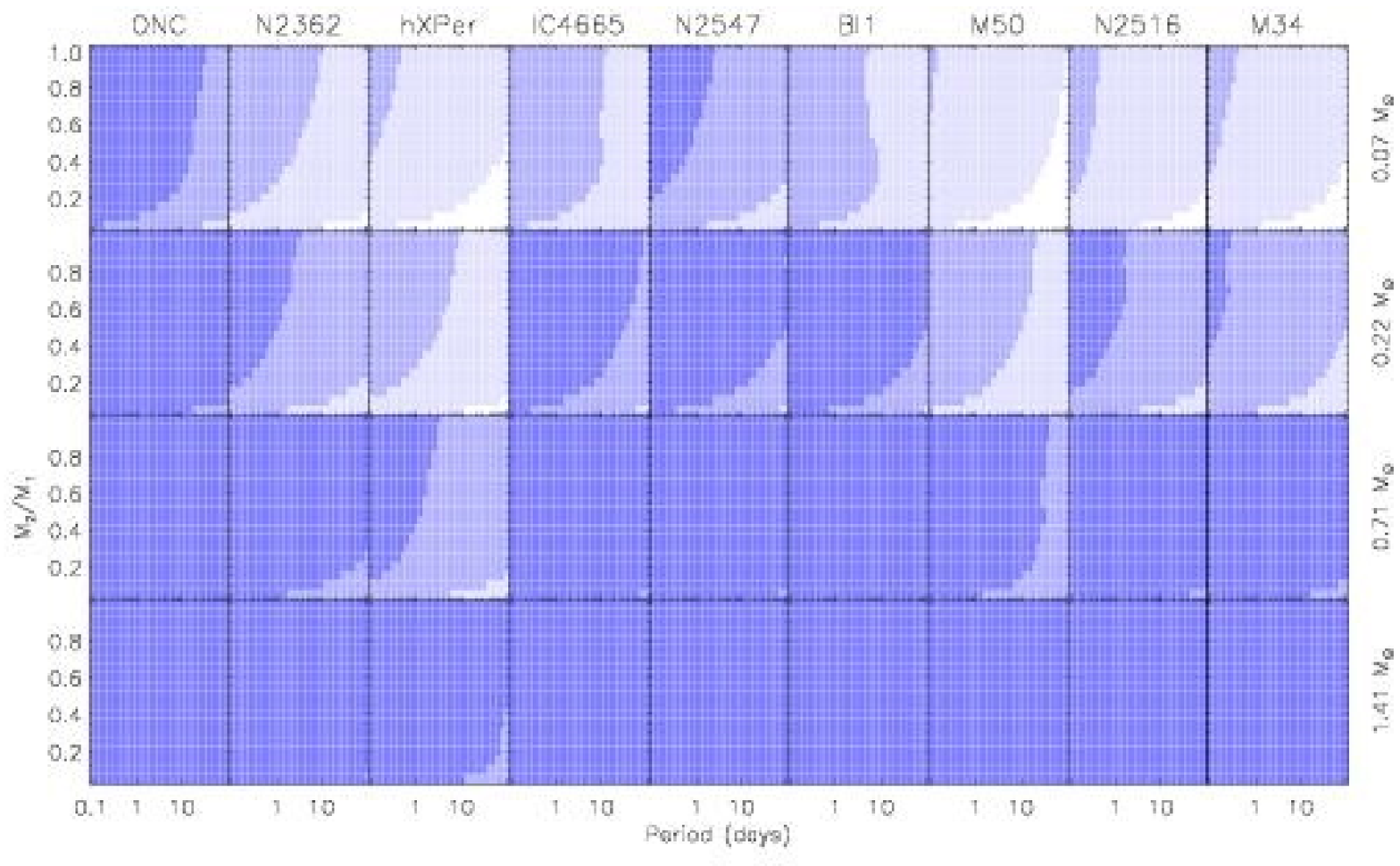,width=\linewidth}
  \end{center}
  \caption{Estimated sensitivity to the RV modulations induced by
    stellar and substellar companions as a function of orbital period
    (x-axis) and mass ratio (y-axis) for each cluster (columns) and
    selected total system masses (rows). Dark, medium and light
    shading correspond to areas where M850, H850 and H600 observations
    respectively should allow us to detect the RV
    modulation. \label{fig:pdet_rv_eb}}
\end{figure*} 

The example RV semi amplitudes shown in the top right panel of
Figure~\ref{fig:rvprec} show that we should be able to detect the RV
modulation induced in the primary of most stellar and substellar
binaries in our survey with medium resolution instruments, while a
small fraction will require 8\,m class follow-up. Only H600
observations of the brightest, slowly rotating stars allow the
detection of planetary companions. This is illustrated on a cluster-by
cluster basis in Figure~\ref{fig:pdet_rv_eb}, where we have shaded the
areas of parameter space for which the RV modulations are detectable
with each type of spectroscopic observations considered. We considered
detectable any system where the RV semi amplitude $K$ is more than
twice the estimated RV precision, i.e. well-timed observations should
enable the detection of the RV modulation at the $4\sigma$
level. Comparison with Figure~\ref{fig:pdet} highlights the good
overlap between the photometric and RV sensitivity for binaries.

Table~\ref{tab:pdet_rv} shows, for each of the types of spectroscopic
observations considered, the percentage of detected eclipses for which
we also expect to detect the RV modulations. We have used the best
case scenario, i.e. zero rotation and most appropriate instrument for
the magnitude and rotation rate considered. 

\begin{table}
  \begin{minipage}{\linewidth}
    \caption{Percentage of true eclipse / transit candidates with
      detectable radial velocity modulations. The letter a, b and c
      refer to the type of RV observations (see text). The last column
      gives the expected number of transiting planets whose RV
      modulation should be detectable in each cluster.}
    \label{tab:pdet_rv}
    \begin{center}
      \begin{tabular}{lrrrrr}
        \hline
        Name     & \multicolumn{3}{c}{Binaries} & \multicolumn{2}{c}{Planets} \\
        & a & b & c & \multicolumn{2}{c}{c} \\
        & \% & \% & \% & \% & No.\ \\
        \hline
        ONC & 84 & 100 & 100 & 100 & 2.3 \\
        NGC\,2362 & 48 & 99 & 100 & 0 & 0.0 \\
        $h$ \& $\chi$\,Per & 24 & 93 & 100 & 0 & 0.0 \\
        IC\,4665 & 69 & 98 & 100 & 20 & 0.6 \\
        NGC\,2547 & 76 & 100 & 100 & 28 & 0.3 \\
        Blanco\,1 & 89 & 100 & 100 & 10 & 0.0 \\
        M50 & 48 & 98 & 100 & 10 & 0.1 \\
        NGC\,2516 & 74 & 99 & 100 & 0 & 0.0 \\
        M34 & 59 & 98 & 100 & 0 & 0.0 \\ 
        \hline
        Total & 45 & 96 & 100 & 27 & 2.8 \\
        \hline
      \end{tabular}
    \end{center}
  \end{minipage}
\end{table}

Except for $h$ \& $\chi$\,Per and M50, where we are primarily monitoring
relatively massive primaries, the vast majority of the candidate
eclipsing binaries will cause RV modulations detectable with M850
observations, i.e.\ after the first stage of our follow-up
observation. Virtually all binary systems in all clusters that are
detectable with photometry are also detectable in RV with H850
observations, and a fortiori with H600 observations. The numbers
involved in $h$ \& $\chi$\,Per are so large that it is unrealistic to
expect all of the candidates to be followed-up. Efficient follow-up of
candidates in these twin clusters will require a Northern hemisphere
high-resolution spectrograph with wide-field multiplexing capabilities
such as WFMOS on SUBARU \citep{bn05}.

For planets, H600 observations only have the potential to detect RV
modulations. In the last two columns of Table~\ref{tab:pdet_rv}, we
give the percentage and number of transiting planets detected
photometrically whose RV modulations should also be detectable with
this type of observation, assuming a planet mass of $M_{\rm
  pl}=1\,M_{\rm Jup}$ for $M_{\rm pl}\geq 0.7\,M_{\rm Jup}$ and
$M_{\rm pl}=0.3\,M_{\rm Jup}$ for $R_{\rm pl}<0.7\,R_{\rm Jup}$. For
Jupiter-mass planets, the feasibility of RV follow-up is essentially
dependent on the cluster distance, and as such will be particularly
problematic in NGC\,2362 (transit detection probabilities are also low
in this cluster), $h$ \& $\chi$\,Per and M50, and to a lesser extent
in M34. The majority of the Jupiter-mass planets causing detectable
transits in the other clusters should induce RV modulations detectable
with instruments such as UVES in the $V$-band, provided their parent
stars are not rotating too fast (in particular, a significant fraction
of planet host-stars in the ONC may be rapid rotators). The RV
modulations induced by low-mass planets are very hard to detect in any
of the target clusters. The complete end-to-end simulations predict
around 3 confirmed detections overall, but one should bear in mind
that this number may go up or down by a factor of two or more if one
tunes the input assumptions within a reasonable range.

\begin{table}
  \begin{minipage}{\linewidth}
    \caption{Number of planetary systems detectable both via their
      transits and in radial velocity if every star with mass above
      $0.2\,M_{\sun}$ possesses a Jupiter mass companion in the range
      0.4--10\,d assuming a the `standard' period distribution, i.e.\ 5 times as many Hot Jupiters as very Hot
      Jupiters (1), or a period distribution that is uniform in $\log P$ (2)..}
    \label{tab:pl_rv}
    \begin{center}
      \begin{tabular}{lrr}
        \hline
        Name     & \multicolumn{2}{c}{Number} \\
                 & (1) & (2) \\
        \hline
        ONC & 139.4 & 1917 \\
        NGC\,2362 & 4.2 & 127 \\
        $h$ \& $\chi$\,Per & 2.5 & 74 \\
        IC\,4665 & 21.0 & 341 \\
        NGC\,2547 & 15.4 & 299 \\
        Blanco\,1 & 6.7 & 105 \\
        M50 & 6.1 & 183 \\
        NGC\,2516 & 7.8 & 235\\
        M34 & 6.8 & 203 \\
        \hline
        Total & 209.8 & 3482\\
        \hline
      \end{tabular}
    \end{center}
  \end{minipage}
\end{table}

It is interesting to examine the results under a different, highly
optimistic set of assumptions regarding planetary companion incidence,
to see whether Monitor will be able to place any kind of constraints
on this incidence. Table~\ref{tab:pl_rv} lists the number of
confirmable detections expected if every star with mass above
$0.2\,M_{\sun}$, hosts a Hot or very Hot Jupiter. This is by no means
a realistic scenario: although the actual incidence of Hot and very
Hot Jupiters may well be higher around very young stars than at later
stages, for example if a `survival of the lucky few' scenario applies,
where most planets migrate into their parent stars and only those that
form shortly before the disk disappears survive, this would at most
imply a factor of a few increase in the incidence of planetary
companions at early ages. However, it serves to illustrate that, if we
do not detect planets in a given cluster, this will imply a strong
upper limit on the incidence of close-in Jupiters in that cluster. Two
possibilities for the period distribution were investigated, namely
the same period distribution as used previsously, where Hot planets
are five times more abundant than very Hot planets, and a distribution
that is uniform in $\log P$. There is an increase by a factor $>10$ in
the numbers for the second case, which illustrates the very
strong bias of the detection method used (transits and radial
velocities combined) towards short periods.

In practice, we will not know a priori which events are transits and
which are eclipses. In cases where the occultation depth and duration
in the initial light curve are consistent with a very low-mass
occulting companion but no RV variations are detected, we will have to
give careful consideration to what phenomena could mimic a planetary
signal in our light curves without generating a detectable RV
signal. Provided that cluster membership can be reliably assessed, a
number of types of `difficult mimics' remain: physical triple systems
belonging to the cluster, starspots, and occultations by warps or
accretion columns in nearly edge-on circumstellar disks. Simultaneous
multi-band monitoring, including the full visible range and the
near-IR, may help discriminate between these and planetary companions,
but we cannot exclude the possibility that low mass brown dwarfs or
planets whose transits we might detect will remain unconfirmed given
the capabilities of present-day instrumentation. Even if every other
hypothesis were excluded, detection of a planetary companion without a
mass estimate would be of limited use in constraining formation and
evolution models. We point out that these `unsolved systems' will make
interesting targets for extremely large telescopes (ELTs) foreseen to
come into operation in the 2010--2020 period, such as the European ELT
or the Thirty Meter Telescope (TMT), assuming they will be equipped
with high resolution visible and near-IR spectroscopic instruments.

\section{Conclusions and future prospects}
\label{sec:status}

We have undertaken a unprecedented survey whose primary goal is to
search for occultations in all suitably young, nearby and
rich young clusters with well-characterised pre-main sequence
populations.

In this paper, we detailed the motivations for undertaking such a
survey, highlighting the fact that occulting companions to young,
low-mass stars constitute a critical area of parameter space, which
has not been explored so far, and where each detection has the
potential to act as a vital anchor point for formation and
evolutionary models of low mass stars, brown dwarfs and planets.

After laying down the considerations which guided the design of the
survey, many of which are dictated by availability of suitable
instrumentation and other circumstances, rather than fully controlled,
we have performed a detailed \emph{a priori} assessment of the
expected performance of the survey in each cluster and as a whole,
incorporating the actual (observed) noise budget and time sampling of
the observations wherever possible, and taking into account the limits
imposed by the need to follow-up candidates spectroscopically as well
as the detectability of occultations.

This has allowed us to explore which area of occulting systems
parameter space we expect Monitor to be sensitive to, and what the
limiting factors are. The range of total system masses which we
observe with useful photometric precision depends on the cluster age
and distance, the telescope aperture and the exposure time, but spans
the entire low-mass star regime, from $1.4\,M_{\sun}$ to the Hydrogen
burning mass limit (HBML), if one considers the entire target
sample. We are primarily sensitive to short-period systems, but this
bias is somewhat reduced by the fact that many occultations of
interest are deep and hence clearly detectable even if a single event
is observed (though a requirement that several events be observed was
imposed when evaluating the detection rates).

Using a set of baseline assumptions for the incidence and parameter
distribution of stellar and substellar companions, we have estimated
the number of eclipsing binary systems that Monitor as a whole should
detect. We find that Monitor should detect approximately 114 such
systems. Close to half of the expected detections are in the twin
clusters $h$ \& $\chi$\,Per, which our preliminary membership study
confirms as extremely rich, and where we are mainly sensitive to total
masses close to $1\,M_{\sun}$. The radial velocity modulations of
these systems are detectable from 8\,m class telescopes, though to
follow-up all the expected candidates would require a very large
allocation of telescope time. Monitor will thus have a very
significant impact in constraining the multiplicity and early
evolution of very low-mass stars and brown dwarfs, bringing either an
increase of several hundred percent on the number of such systems
known or extremely stringent constraints on their incidence in the
event of non-detections.

Transits by Hot and Very Hot Jupiters are also detectable in all
clusters, in some cases only around solar mass stars but in others
down to the HBML and below (if such planets exist). Additionally, we
concur with the prediction of \citet{pg05b} that transits by
relatively small ($< 0.5\,R_{\rm Jup}$ planets are detectable in
nearby, young (few tens of Myr) clusters, except at very early ages
($\leq 13$\,Myr) where the stars are too large for transits of
sub-Jupiter radius to be detected. Under assumptions regarding the
incidence of planetary companions which are compatible with current
observational evidence, we foresee the detection of transits of few
planets in the ONC and IC\,4665, with around 0.5--1 detections in the
other clusters. Around 30\% of these, mostly in the ONC, should induce
radial velocity variations in their parent stars that are detectable
with optical high-resolution spectrographs on 8\,m class telescopes
provided the stars are not rapid rotators, giving a final estimate of
$\sim 3$ confirmable transiting planet detections. Therefore, the
Monitor project is well placed to detect the first transiting
extra-solar planet(s) in young open cluster(s). Even a single
detection in any cluster will be significant, as it would constrain a
completely new region of the age-mass-radius relation. Additionally,
non detections of any transits (i.e.\ if all candidates in a given
cluster turn out to be contaminants or binaries) in a given cluster
would place strong constraints on the incidence of short-period
planets in that cluster.

We intend to use the results presented here as a benchmark against
which to compare the actual results of the survey in each cluster. At
the time of writing, photometric monitoring is complete for 1/3 of the
target clusters. The light curves for these clusters are under
analysis and 37 high-quality candidates with CMD positions compatible
with cluster membership and at least 3 observed occultations, have
been identified so far (4, 12, 20 and 1 in the fields of the ONC,
NGC\,2362, M50 and M34 respectively). Given that this census is both
incomplete, because final refinements to the light curve
pre-processing and transit search procedure still remain to be made
and only candidates brighter than $I=18$ were searched for so far, and
contaminated by field systems, the numbers are broadly consistent with
the values in Table~\ref{tab:ndet}. The first radial velocity
observations we obtained for some of these candidates in late 2005 --
early 2006, and full-scale radial velocity follow-up observations will
start in earnest in the winter 2006--2007 observing season. A more
detailed comparison will be carried out when the eclipse search is
complete and foreground and background contaminants have been
identified spectroscopically.

Aside from spectroscopic follow-up of individual objects, two
extensions of the Monitor project are foreseen to complement the main,
optical monitoring survey. The first extension consists of photometric
monitoring in the near-infrared, using new wide-field facilities such
as WIRCAM on the CFHT and WFCAM on UKIRT. This will provide increased
sensitivity to occultations of low-mass objects and to secondary
occultations, as well as enabling us to characterise stellar
variability in more detail. Snapshot-mode near-IR monitoring of the
ONC using WIRCAM on the CFHT is due to start in 2006B. The second
extension consists of spectroscopy of a large number of candidate
members in each cluster, using wide-field multi-fibre visible and
near-IR spectrographs. This will provide robust membership catalogs,
enabling us to study accretion and lithium depletion. Wherever possible,
this membership survey will be combined with the radial velocity
follow-up of occultation candidates and carried out over multiple
epochs, allowing us to search for spectroscopic binaries. The first of
these multi-fibre surveys, targeting the ONC and M34, are scheduled
for the fall of 2006 using FLAMES on the VLT and WYFFOS on the WHT
respectively.

\section*{Acknowledgments}

The authors wish to thank Jerome Bouvier, Cathie Clarke, Ettore
Flaccomio and Mark McCaughrean for useful discussions,
and Dan Bramich, Richard Alexander and Patricia Verrier for help with
observing. SA gratefully acknowledges support from a PPARC
Postdoctoral Research Fellowship and JMI from a PPARC studentship.

The Isaac Newton Telescope is operated on the island of La Palma by
the Isaac Newton Group in the Spanish Observatorio del Roque de los
Muchachos of the Instituto de Astrofisica de Canarias. Based on
observations made with ESO Telescopes at the La Silla Observatory
(program ID 175.C-0685). Based on observations obtained with
MegaPrime/MegaCam, a joint project of CFHT and CEA/DAPNIA, at the
Canada-France-Hawaii Telescope (CFHT) which is operated by the
National Research Council (NRC) of Canada, the Institute National des
Sciences de l'Univers of the Centre National de la Recherche
Scientifique of France, and the University of Hawaii. Based on
observations obtained at Cerro Tololo Inter-American Observatory and
Kitt Peak National Observatory, National Optical Astronomy
Observatories, which are operated by the Association of Universities
for Research in Astronomy, under contract with the National Science
Foundation.

This research has made use of the WEBDA Database,
developed by J.-C. Mermilliod at the Laboratory of Astrophysics of
the Ecole Polytechnique F{\' e}d{\' e}rale de Lausanne and maintained
by Ernst Paunzen at the Institute of Astronomy of the University of
Vienna; and of the VALD database, operated at the Institute for
Astronomy of the University of Vienna.

We are also grateful to the referee, Scott Gaudi, for his careful
reading and his constructive comments, which helped make this a better
paper.

\bibliographystyle{mn2e} \bibliography{aigrain}

\appendix

\section{RV precision estimates}
\label{app:rvprec}

\subsection{M\&H850}

We also investigated the RV accuracy achievable using the same
instruments based on measuring the positions of the three Ca~T
lines. When using a small number of lines, radial velocities are
derived by fitting the profiles of individual lines and taking a
(weighted) average of the results from all the lines. The observed
profile of each line is the convolution of its intrinsic width with a
number of broadening processes -- including pressure broadening,
thermal broadening, micro- and macro-turbulence and rotation -- and
the instrumental profile. The Ca~T lines are saturated and therefore
intrinsically broad, and this intrinsic width dominates for
slowly-rotating stars. We have measured the width of the Ca~T lines
in previous FLAMES$+$GIRAFFE (low resolution mode) spectra of
slowly-rotating M-type stars (both dwarfs and giants) to be
$\delta\lambda\sim 0.3$\,nm. Other important contributions to the line
width for young M-type stars are macro-turbulence ($\xi \sim 1$\,km/s,
\citealt{gra05}, Appendix B) and rotation (expected projected
rotational velocities for our stars range from $v \sin i \sim 5$ to
100\,km/s), the natural width and thermal and pressure broadening
being negligible in comparison\footnote{Natural and pressure
  broadening widths were checked using the Vienna Atomic Line Database
  \citep{kpr+99}.}.

While the intrinsic line profile has a complex shape with a flat core
and broad wings (for some young objects, emission is seen in the
core), both rotation and macro turbulence give rise to Gaussian
profiles, and the instrumental profile is also well approximated by a
Gaussian. The measurement of radial velocities is thus well
approximated by a Gaussian fitting process. From $\chi^2$ minimisation
with respect to the line centre, the precision with which one can
measure the RV from each line is given by:
\begin{equation}
\label{eq:rvsig_fit}
\sigma_{\rm RV} ({\rm line}) = 
\frac{c}{\lambda}~\frac{\Delta\lambda~\left( 
 4 \pi^{1/2}~\Delta\lambda \right)^{1/2}}
                       {S~{\rm EW}},
\end{equation}
where $c$ is the speed of light, $\lambda$ is the line wavelength,
$\Delta\lambda$ the observed full width at half-maximum of the line
(in wavelength units), $S$ is the signal-to-noise per wavelength unit
in the continuum, and {\rm EW} is the line equivalent width. In the
Sun, the EWs of the Ca~T lines at 849.8, 854.2 and 866.2\,nm are
0.146, 0.367 and 0.260\,nm respectively, (\citealt{gra05}, Appendix
E). We model $\Delta\lambda$ as:
\begin{equation}
\label{eq:fwhm}
\left( \Delta \lambda \right)^2
           = \left( \delta\lambda^2 \right)^2 + 
             \left(\frac{\lambda}{R} \right)^2 + 
             \left( v \sin i \right)^2 + \xi^2,
\end{equation}
where $R$ is the resolution of the spectrograph.
Combining the RV measurements from the three individual Ca~T
lines gives:
\begin{equation}
\label{eq:rv_3lines}
\sigma_{\rm rv} ({\rm combined}) = 
  \left( \sum_{\rm lines} \frac{1}{\sigma^2_{\rm rv} ({\rm line})} \right)^{-1/2},
\end{equation}

The ThAr lamp cannot be used at the wavelength of the Ca~T because
very bright Ar lines saturate the detector and contaminate nearby
spectra. The wavelength calibration is thus based on sky emission
lines. Our tests with sky spectra extracted from the ESO archive to
which noise was artificially added show that the systematic wavelength
calibration errors based on sky lines are below 200\,m/s with GIRAFFE
and UVES (both modes).

\subsection{H600}

We can use the experience of the OGLE follow-up campaigns to evaluate
the RV accuracy achievable with spectra taken in the 600\,nm region
with FLAMES$+$GIRAFFE, FLAMES$+$UVES or UVES in slit mode, with
simultaneous wavelength calibration, based on the following
information:
\begin{itemize}
\item The precision $\sigma_{\rm RV}$ with which one can measure the
  RV scales linearly with the signal-to-noise ratio per pixel in the
  continuum.
\item With FLAMES$+$UVES at $I=15$ (for objects with $V-I \sim 1.5$),
  one can achieve $\sigma_{\rm RV}=0.05$, 0.1, 1 and 3\,km/s for $v
  \sin i=0$, 20, 40 and 60\,km/s respectively, in 1\,h
  exposures. Additionally, a residual wavelength calibration error of
  35\,m/s (in good atmospheric conditions) must be added in quadrature
  to $\sigma_{\rm RV}$.
\item Relative to the multi-fibre mode, using UVES in slit mode
  results in a gain of a factor of 3 in signal-to-noise ratio, and
  hence in $\sigma_{\rm RV}$, at the cost of observing only one object
  at a time and of an additional systematic error component of
  150\,m/s resulting from the seeing.
\item For the same magnitude, colour and exposure time, $\sigma_{\rm
    RV}=60$\,m/s for $v \sin i=0$\,km/s with FLAMES$+$GIRAFFE at
  $I=15$. We shall assume that $\sigma_{\rm RV}$ scales with $v \sin
  i$ in the same way with FLAMES$+$GIRAFFE as it does with
  FLAMES$+$UVES.
\end{itemize}

\subsection{General considerations}

High activity levels are associated with surface convective
inhomogeneities and starspots, which induce a RV `jitter'. Based on
long term monitoring of a large number of field stars, \citet{sbm98}
showed that the activity-induced jitter is essentially proportional to
the projected rotational velocity $v \sin i$ for G and K stars, and
\citet{psc+02} confirmed this trend for Hyades stars. Extrapolating
their relations to early ages, we expect it to reach up to $30$\,m/s
at the age of M34, and $\sim100$\,m/s at the age of the ONC. In both
sets of calculations ($V$-band and Ca~T), we assumed 50\,m/s RV
jitter, added in quadrature to the other components. 

The signal-to-noise as a function of $I$-band magnitude was estimated
for each instrument using the ESO Exposure Time
Calculators\footnote{Available from {\tt
    http://www.eso.org/observing/etc/}.}, using a \citet{pic98} M2V
template spectrum and assuming the following observing conditions:
seeing $\leq 0.8$\arcsec, airmass $\leq 1.6$, 3 days from new moon, and
fibre positioning errors $\leq 0.1$\arcsec. The following
observational setups were used:
\begin{itemize}
\item{\bf M850 :} EMMI on the NTT (grating 9, cross-disperser CD4,
  central wavelength 850\,nm, 1\arcsec slit), taking into account the
  fact that we will tailor exposure times to the apparent magnitude of
  each object (up to 1\,h) so as to ensure an overall limiting
  accuracy of approximately 1.5\,km/s.
\item{\bf H850:} FLAMES$+$GIRAFFE setup H21, FLAMES$+$UVES
  standard setup with cross-disperser CD4 and central wavelength
  860\,nm, and same setup for UVES in slit mode, assuming a 0.8\arcsec
  slit. Note that with this UVES standard setup, the strongest of the
  three Ca~T lines falls in the gap between the two CCDs. This was
  not taken into account in the present calculations, the goal being
  to test whether there was any case for a modified standard setup
  avoiding this drawback.
\item{\bf H600:} FLAMES$+$GIRAFFE setting H15n, FLAMES$+$UVES
  standard setup with cross-disperser CD4 and central wavelength
  580\,nm, and same setup for UVES in slit mode, assuming a 0.8\arcsec
  slit.
\end{itemize}
\bsp

\label{lastpage}

\end{document}